\documentclass[pra,aps,twocolumn,showpacs,superscriptaddress,nofootinbib]{revtex4-1}
\usepackage{graphicx}
\usepackage{dcolumn}
\usepackage{bm}
\usepackage[bbgreekl]{mathbbol}
\usepackage{amsmath}
\def\Eb{{\bf E}}
\def\g2R{g^{(2)}_{\rm R}}

\def\w0{w_0}
\def\avg#1{\mathinner{\langle{#1}\rangle}}
\usepackage[dvipsnames]{xcolor} 

\def\ii{{\rm i}}  
\def\GG{{\bf G}}

\def\db{{\bf d}}  
\def\Eb{{\bf E}}  
\def\rb{{\bf r}}

\def\hrr{\hat{\sigma}^{rr}}  
\def\db{\textbf{d}}

\def\ket#1{\mathinner{|{#1}\rangle}}
\def\db{\boldsymbol{\wp}}

\def\db{\boldsymbol{\wp}}  
 
\def\hEb{\hat{E}}
\def\heg{\hat{\sigma}^{eg}}  
\def\hge{\hat{\sigma}^{ge}}  
\def\hee{\hat{\sigma}^{ee}}

\def\Kmax{K^{\rm max}}

\def\Er{\hEb_{\rm R}}

\def\Etin{\hEb_{\rm T, in}}

\def\Heff{\mathcal{H}_{\rm eff}}
\def\hrho{\hat{\rho}}
\def\kpar{{\bf k}_{\parallel}}

\usepackage{color}

\begin{document}
\title{Quantum nonlinear optics based on two-dimensional Rydberg atom arrays}
\author{M. Moreno-Cardoner}
\thanks{These authors have equally contributed to this work.}
\address{Institut f\"ur Theoretische Physik, Universit\"at Innsbruck, Technikerstr. 21a, A-6020 Innsbruck, Austria.}
\address{Departament de F{\'i}sica Qu{\`a}ntica i Astrof{\'i}sica and Institut de Ci{\`e}ncies del Cosmos, Universitat de Barcelona, Mart{\'i} i Franqu{\`e}s 1, E-08028 Barcelona, Spain.}
\author{D. Goncalves}
\thanks{These authors have equally contributed to this work.}
\address{ICFO-Institut de Ciencies Fotoniques, The Barcelona Institute of Science and Technology, 08860 Castelldefels (Barcelona), Spain.}
\author{D. E. Chang}
\address{ICFO-Institut de Ciencies Fotoniques, The Barcelona Institute of Science and Technology, 08860 Castelldefels (Barcelona), Spain.}
\address{ICREA-Instituci{\'o} Catalana de Recerca i Estudis Avan{\c c}ats, 08015 Barcelona, Spain.}

\date{\today}

\begin{abstract}
\vspace{1em}
Here, we explore the combination of sub-wavelength, two-dimensional atomic arrays and Rydberg interactions as a powerful platform to realize strong, coherent interactions between individual photons with high fidelity. In particular, the spatial ordering of the atoms guarantees efficient atom-light interactions without the possibility of scattering light into unwanted directions, for example, allowing the array to act as a perfect mirror for individual photons. In turn, Rydberg interactions enable single photons to alter the optical response of the array within a potentially large blockade radius $R_b$, which can effectively punch a large ``hole'' for subsequent photons. We show that such a system enables a coherent photon-photon gate or switch, with an error scaling $\sim R_b^{-4}$ that is significantly better than the best known scaling in a disordered ensemble. We also investigate the optical properties of the system in the limit of strong input intensities. Although this a priori represents a complicated, many-body quantum driven dissipative system, we find that the behavior can be captured well by a semi-classical model based on holes punched in a classical mirror. 
\end{abstract}
\maketitle

\textit{Introduction.--}Achieving strong and controlled optical nonlinearities at the level of single photons represents one of the greatest goals and challenges within quantum and nonlinear optics~\cite{ChVL14}. A highly promising approach in recent years has emerged based upon Rydberg electromagnetically induced transparency (rEIT). In standard EIT, an additional pump field enables probe photons to hybridize with meta-stable atomic excitations and propagate without loss (see Fig. 1a,b)~\cite{FIM05,PFM01,LDB01}. When the metastable state corresponds to a high-lying Rydberg level (rEIT), this effect becomes highly nonlinear as strong atomic Rydberg interactions destroy the resonance condition needed for EIT. Then, a second photon within a ``blockade radius'' of the first effectively sees a highly scattering two-level medium. Such strong photon-photon interactions mediated by Rydberg atoms are now routinely observed in experiments~\cite{FPL13,BTR14,TNL17,ref:GateRempe19,ref:CSAdams10,ref:PRL113}. 

However, despite many spectacular experiments, it remains challenging to functionalize rEIT into coherent, single-photon-level nonlinear devices. A major reason is that the two-level atomic medium within the blockade radius is naturally dissipative, and tends to scatter the second photon into random uncontrolled directions. Dissipation can be suppressed, but this also reduces the coherent response and necessitates a large resource overhead to compensate~\cite{ref:Gorshkov11}. The best known gate protocol has a theoretically predicted error that scales with blockade radius (or more properly, optical depth per blockade radius) as $\sim R_b^{-3/2}$~\cite{TNL17}.

\begin{figure}
    \centering
\begin{minipage}{0.45\textwidth}
    \centering
    \includegraphics[width=\textwidth]{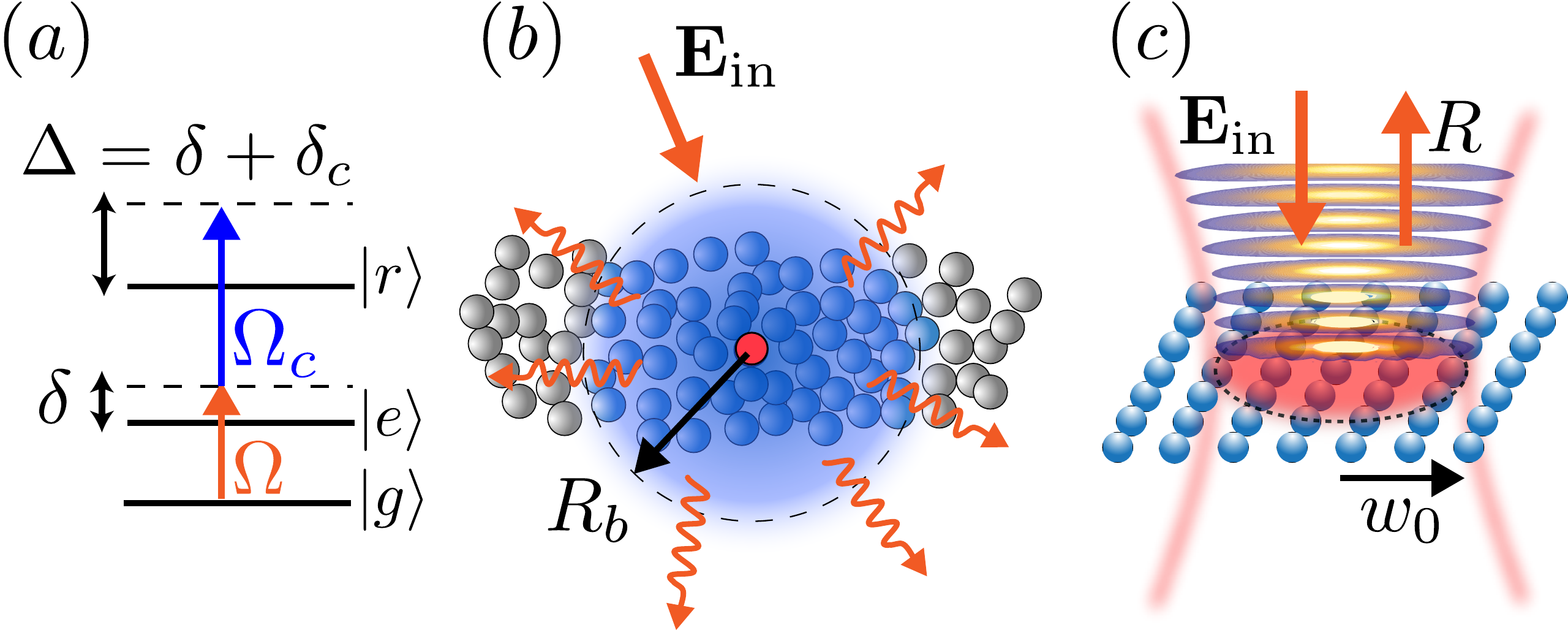}
    \end{minipage}
        \caption{(a) Level diagram of single three-level atom, with ground ($|g\rangle$), excited~($|e\rangle$), and Rydberg~($|r\rangle$) levels, and relevant detunings and Rabi frequencies indicated. (b) In rEIT in a disordered ensemble, an atom in a Rydberg level~(red) creates an effective medium of two-level atoms~(blue) within a blockade radius $R_b$, which strongly scatters near-resonant, incident light.  (c) An array of two-level atoms (with states $|g,e\rangle$) perfectly reflects weak resonant light. Here, we illustrate a Gaussian beam with waist $w_0$.}
    \label{Fig1}
\end{figure}

Intuitively, a much more robust path to quantum nonlinear optics could be established, if an ensemble of two-level atoms could be made completely lossless, even for resonant light. Remarkably, this can occur when the atoms are positioned in a defect-free array with sub-wavelength lattice constant. Then, interference in emission combined with the spatial ordering ensures that atoms cannot scatter light into random directions, but only into the same mode (either in the backward or forward directions) as the light coming in. The optical properties of arrays have thus attracted significant interest, especially in the linear optical regime \cite{ZR11,JR12,POR15,BGA15,BGA16,BGA16b,SR16,SWL17,PBCh17,MMA18,PR19,RWR20,ZM19,ZChM20,PR20,ref:QuantumMetasurfaces,ref:Cirac}. As one particularly relevant example, it has been theoretically predicted \cite{DeA07,BGA16b,SWL17} and experimentally observed~\cite{RWR20} that a two-dimensional~(2D) array can act as a nearly perfect mirror for weak resonant light~(Fig.~\ref{Fig1}c). The nonlinear optical properties of arrays have also begun to be explored, such as using the two-level nature of the atoms themselves~\cite{ref:Bettles2020,ref:Subwavelength1,ref:Subwavelength2} or the atomic motion~\cite{SLY20}, as has the conditioning of the linear response, based on Rydberg blockade, to produce interesting quantum states of light~\cite{ref:QuantumMetasurfaces,ref:Cirac}.

Here, we explore the combination of 2D arrays and Rydberg interactions as a powerful platform for quantum nonlinear optics. First, we investigate the second-order quantum correlations of the reflected field, in the presence of Rydberg dressing interactions. In particular, we show that the field becomes strongly anti-bunched once the blockade radius exceeds the incident beam waist, indicating the inability of the system to simultaneously reflect two photons at once. This intuitively results from a reflected photon momentarily punching a large ``hole'' in the atomic mirror. Motivated by this signature of strong nonlinearities, we then propose a protocol to realize a coherent photon-photon gate or switch, where the presence (absence) of a first photon results in the transmission (reflection) of a second photon. We also show that our approach provides a gate error with a very favorable scaling, decreasing with blockade radius as $\sim R_b^{-4}$. Finally, we investigate the optical response of such an array in the many-body limit of high input intensity. We find that the system exhibits a non-trivial dependence of reflectance, transmittance, and loss on driving power and blockade radius. Interestingly, although this a priori represents a complicated, many-body quantum driven dissipative system, we find that the behavior can be captured well by a semi-classical stochastic model based on holes punched in the mirror. \\

\textit{System and Formalism.--} Let us consider a two-dimensional square array (lattice constant $d$) of $N_a = N^2$ atoms trapped at fixed positions in free space in the $z=0$ plane, and with ground and excited states ($|g\rangle$ and $|e\rangle$ respectively) with an optically allowed transition with electric dipole matrix element $\boldsymbol{\wp}$ oriented along one of the lattice axes. Each atomic transition will not only interact with an incoming field, which we take to be a coherent state with spatial mode $\Eb_{\rm in}(\rb)$ and frequency $\omega_L$, but also the fields produced by other atoms. Starting from the full atom-light Hamiltonian and integrating out the photonic degrees of freedom within the Born-Markov approximation~\cite{MS1990}, the atoms are governed by the master equation 
\begin{subequations}\label{Eq:dynamics}
\begin{equation}
\dot{\hat{\rho}} = -(\ii/\hbar) \left(\Heff\hat{\rho}-\hat{\rho}\Heff^{\dagger} \right)+\mathcal{L}_{\rm jump} \left[ \hat{\rho} \right]\label{Eq:master},
\end{equation}
\begin{multline}
\mathcal{\Heff}/\hbar=-\left(\delta+i\frac{\Gamma_0}{2}\right)\, \sum_{i=1}^{N_a}\hee_i - \sum_{i=1}^{N_a}\left(\Omega_i \hge_i + h.c.\right) \\
+ \sum_{i,j=1, i\neq j}^{N_a} \left(J^{ij}-i \frac{\Gamma^{ij}}{2}\right) \heg_i \hge_j +\hat{V}_{\rm Ryd}\label{Eq:Heff},
\end{multline}
\begin{equation} 
\mathcal{L}_{\rm jump}[\hrho]=\sum_{i,j=1}^{N_a}\Gamma^{ij}\,\hge_j\hrho\heg_i.
\label{Eq:Jumps}
\end{equation}
\end{subequations}
Here we define the atomic operators $\hat{\sigma}_i^{\alpha\beta}=|\alpha_i\rangle\langle\beta_i|$ with $\{\alpha,\beta\}\in \{g,e\}$, $\delta =\omega_L-\omega_0$ is the detuning with respect to the single atom bare frequency $\omega_0$, and $\Omega_i = \boldsymbol{\wp} \cdot \Eb_{\rm in} (\rb_i)$ is the Rabi frequency at atomic position $\rb_i$. The term $\hat{V}_{\rm Ryd}$ associated with Rydberg interactions will be specified later. The photon-mediated dipole-dipole interactions between atoms are characterized by
\begin{subequations}\label{Eq:rates}
\begin{align}\label{shiftrate}
J^{ij}&=-\frac{\mu_0\omega_0^2}{\hbar}\,\db^*\cdot\text{Re}\,\mathbf{G}(\rb_i-\rb_j,\omega_0)\cdot\db, 
\end{align}
\begin{equation}\label{dissiprate}
\Gamma^{ij} =\frac{2\mu_0\,\omega_0^2}{\hbar}\,\db^*\cdot\text{Im}\,\mathbf{G}(\rb_i-\rb_j,\omega_0)\cdot\db, 
\end{equation}
\end{subequations}
with $J^{ij}$ and $\Gamma^{ij}$ describing coherent interactions and collective emission, respectively. $\GG (\rb ,\omega_0)$ is the electromagnetic Green's tensor in free space,
\begin{align}
\GG (\rb ,\omega_0) = \frac{e^{i k_0 r}} 
{4\pi k_0^2 r^3} \left[(k_0^2 r^2+ik_0 r -1) \mathbb{1} \right.+ \notag \\
\qquad{}\left. + (-k_0^2 r^2 -3ik_0 r + 3) \frac{\rb \otimes \rb}{r^2}  \right], \label{Greens_function}
\end{align}
with $k_0 = \omega_0/c$. For a single isolated atom, the excited-state spontaneous emission rate is given by $\Gamma^{ii}\equiv\Gamma_0=\wp^2 k_0^ 3/(3\pi \hbar \epsilon_0)$. 

Once the dynamics of Eq.~(\ref{Eq:master}) are solved, the field correlations can be reconstructed from an input-output relation \cite{Gruner1996,Dung2002,Buhmann2007,AHCh17}. Generally, the atoms will emit a quantum field $\hat{\Eb}(\rb)$ into all directions, many of which cannot be detected. Instead, we consider an experimentally realistic scenario, in which an input mode (\textit{e.g.}, a Gaussian) is defined, and light is collected back into the same mode in the backward (reflected) or forward (transmitted) directions. Defining quantum fields projected into these discrete detection modes $\Eb_{\rm det} (\rb)$ (with ${\rm det}=(R,T))$, the input-output relation reads
\begin{equation}
\hEb_{\rm det} = \hEb_{\rm det, in} + i \sqrt{\frac{k_0}{2\hbar\epsilon_0 A}}\sum_{j=1}^{N_a} \Eb^*_{\rm det}(\rb_j) \cdot \db \hge_j
\label{Eq:fields}
\end{equation}
where $\hEb_{\rm det, in}$ is the quantum input field in the particular mode, and $A=\int_{\rm z=const}  \Eb^*_{\rm det}(\rb) \Eb_{\rm det}(\rb) d^{2}\rb$. This normalization is chosen so that $\avg{\hEb^\dagger_{\rm det} \hEb_{\rm det}}$ is the rate of photons detected in the mode.\\

\textit{Linear regime and perfect reflection}.--
We first briefly review a central result to the rest of the paper, that an infinite 2D array can behave as a perfect mirror for single photons. Specifically, let us consider a plane-wave input field $\sim e^{\ii {\bf k}\cdot \rb}$ with in-plane and normal wavevector components $\kpar$ and $k_z$, respectively, satisfying $|\kpar|^2+k_z^2=(\omega_L/c)^2$. We will work in the limit of weak driving field amplitude, where one can restrict the atomic Hilbert space to just a single excitation.  Due to the system periodicity, the driving field only couples to collective spin waves $\ket{\kpar}= N_a^{-1/2}\sum_j e^{\ii \kpar \cdot \rb_j} \heg_j \ket{g}^{\otimes N_a}$, with the same in-plane wavevector. In turn, the periodicity ensures that the spin wave can only emit light in well-defined directions (diffraction orders), with all but the fundamental directions $\pm k_z$ becoming evanescent for sufficiently small lattice constant. Defining the detection modes in Eq. \eqref{Eq:fields} to be transmitted and reflected plane waves ${E}_{R,T}\sim e^{i\kpar\cdot{\bm \rho}\pm i k_z z}$ (with ${\bm \rho}$ being the in-plane position), the reflection and transmission amplitudes in this limit are~\cite{SWL17}
\begin{equation}
r = \frac{\avg{\Er}}{\avg{\Etin}}=-\frac{\ii\Gamma_{\kpar}/2}{\delta-\Delta_{\kpar}+\ii\Gamma_{\kpar}/2}, \qquad{} t = 1+r.
\label{eq:r}
\end{equation}
The condition for only the fundamental diffraction order to be radiative is satisfied for a range of incident wavevectors around ${\bf k}_{\parallel}=0$ when $d<\lambda_0$, and for all incidence angles once $d<\lambda_0/2$. Here, $\Gamma_{\kpar} = (3\pi/k_0^2 d^2) (1- |\kpar|^2/k_0^2)$ and $\Delta_{\kpar} = \sum_{j \neq 0 } e^{\ii \kpar \cdot (\rb_j-\rb_0) } J^{0j}$ are the collective decay rates and resonance frequency shifts (with respect to the bare atomic frequency) of the spin wave mode $\kpar$~\cite{AMA2017}, arising from the interaction of atoms with the fields of other atoms. Notably, when the driving field resonantly excites a spin wave mode~($\delta=\Delta_{\kpar}$), the array becomes purely reflecting, i.e. $|r|^2=1$, as a result of perfect destructive interference in transmission between the incident and re-radiated fields.

In the case of a finite array, the reflection is slightly reduced from unity. For concreteness, we take a Gaussian input beam propagating perpendicularly to the array, whose spatial profile at the atomic plane is $E_{\rm in}({\bf \rho},z=0)=E_0 e^{-\rho^2/w_0^2}$, with beam waist $w_0$~(Fig.~\ref{Fig1}c). Then, the reflectance $R=|r|^2$ is approximately given by \cite{MMA18}
\begin{equation}
R \approx {\rm erf}^4(Nd/\sqrt{2}w_0) -C_R(d)(\lambda/w_0)^4,
\label{eq:reflectance}
\end{equation}
with $C_R(d)$ being a constant depending on the lattice parameter. The first term, containing the error function ${\rm erf}(Nd/\sqrt{2}w_0)$, represents the imperfection of the Gaussian mode extending beyond the array boundaries. The term $C_R(d)(\lambda_0/w_0)^4$ arises from the finite beam waist and persists even for an infinite array. Physically, perfect reflection requires a resonance condition $\delta=\Delta_{\kpar}$ that depends on the incident wavevector, which cannot be simultaneously satisfied for a Gaussian beam that is a superposition of wavevectors. \\

\textit{Rydberg dressing.--} 
We now consider the addition of a high lying Rydberg state $\ket{r}$ coupled to the excited state $\ket{e}$ by a uniform classical control field with Rabi frequency $\Omega_c$, and detuning $\delta_c$ from the bare $\ket{e}$-$\ket{r}$ transition. We also introduce the two-photon detuning $\Delta=\delta+\delta_c$ (see Fig.~\ref{Fig1}a). Rydberg atoms undergo a strong van der Waals interaction, of the form $\hat{V}_{\rm vdW} = \sum_{i\neq j} C_6 r_{ij}^{-6} \hrr_i \hrr_j$, with $r_{ij} = |\rb_i -\rb_j|$.

Instead of the typical rEIT approach for nonlinear optics, we consider an alternative ``Rydberg dressing'' regime, more commonly applied within ultracold atomic physics \cite{ZvBS16,ZChR17}. While rEIT is certainly valid in arrays as well~\cite{ref:QuantumMetasurfaces}, the dressing scheme allows one to make the mirror nonlinear for multiple photons, while simultaneously reducing the Hilbert space from three to two states per atom, which aids calculations in the many-body limit.

We consider two separate dressing schemes, both characterized by a large control field detuning $|\delta_c|\gg \Omega_c$. In general, for just a single atom in state $\ket{e}$, the control field induces an ac-Stark shift $\Delta_{ac}\approx \Omega_c^2/\delta_c-\Omega_c^4/\delta_c^3$, considering corrections up to $\Omega_c^4$. In the first scheme, to be used when calculating second-order correlation functions or in the strong driving limit, the state $|r\rangle$ is only virtually populated, but Rydberg interactions nonetheless disrupt the Stark shift for two nearby atoms in $|e\rangle$, introducing a correction of order $\Omega_c^4/\delta_c^3$. In the second scheme, to be used in the photon gate, an incoming photon will be stored as a single Rydberg excitation, and its presence suppresses the entire ac-Stark shift for any $|e\rangle$ atom within the blockaded region. In either case, the Rydberg dressing interaction can be approximated by a step function,  $\hat{V}_{\rm Ryd}^{ee} \approx  \sum_{i\neq j} V \Theta(R_b - r_{ij}) \hat{\sigma}^{ee}_i \hat{\sigma}^{ee}_j$ or $\hat{V}_{\rm Ryd}^{re} \approx  \sum_{i\neq j} V \Theta(R_b - r_{ij}) \hat{\sigma}^{rr}_i \hat{\sigma}^{ee}_j$ for the first and second schemes, respectively (see Appendix A). The strength of $V$ is only fundamentally limited by laser power, while the blockade radius $R_b$ depends on detuning $\delta_c$ and the $C_6$ coefficient. We will largely work in the simplified limit where $V\rightarrow \infty$, discussing corrections (particularly to the gate fidelity) as relevant. We also apply the convention that $\Theta(0)=1$, e.g., $R_b=d$ implies that nearest neighbors are blockaded.\\

\textit{Optical nonlinearities in the weak driving limit.--} A signature of strong single-photon-level nonlinearities is often revealed by the second-order correlation function, which we now calculate for the reflected field. Specifically, we consider a Gaussian input beam, weakly driving an array of large enough size compared to the beam waist ($w_0/d = 0.35 N$), so that diffraction off the edges can be neglected, and its frequency is aligned with the resonance condition $\delta = \Delta_{{\bf k}_{\parallel} = 0}$. To evaluate the second order correlation in the weak driving limit, we truncate the atomic system at two excitations, solve the steady-state atomic dynamics~(including the Rydberg dressing $\hat{V}^{ee}_{\rm Ryd}$), and then use the input-output relation (\ref{Eq:fields}) to calculate $\g2R =  \avg{\Er^\dagger \Er^\dagger \Er \Er}/\langle\Er^\dagger \Er\rangle^2$ (see Appendix B). Physically, $\g2R$ characterizes the relative probability of immediately detecting a second reflected photon, given the detection of a first reflected photon.

\begin{figure}[t]
\centering
\includegraphics[width=0.45\textwidth]{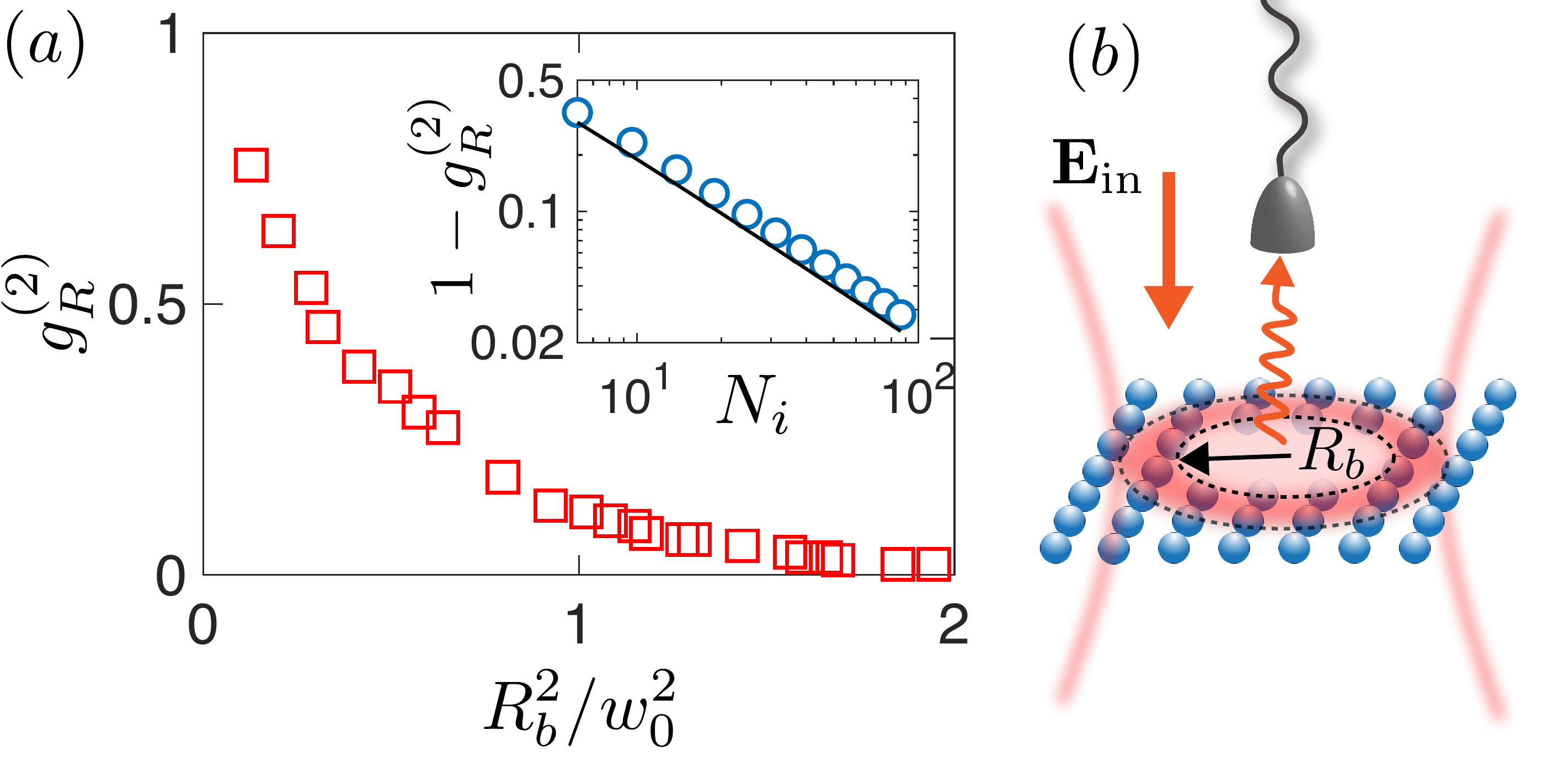}
\caption{(a) Two-photon correlation function in reflection $\g2R$, as a function of the squared ratio of blockade radius to beam waist $R_b^2/w^2_0$, for $N=16$. Strong anti-bunching ($\g2R\rightarrow 0$) is achieved for $R_b > w_0$. The inset shows $\g2R$ in the absence of Rydberg interactions, as a function of approximate number of illuminated atoms $N_i = \pi w_0^2/d^2$ for linear array size $N \in [4,15]$. A small amount of anti-bunching arises due to saturation, as modeled well by $\g2R \sim (1-1/N_i)^2$ (solid line). Other parameters are: beam waist $w_0 = 0.35 Nd$, lattice constant $d/\lambda_0 = 1/2$. (b) After a first photon is detected in reflection a hole of radius $R_b$ is effectively punched in the atomic mirror due to Rydberg blockade.} 
\label{Fig2}
\end{figure}

First, in Fig.~\ref{Fig2}a, we plot the numerical results of $g^{(2)}_R$, as a function of the squared ratio of blockade radius to beam waist $(R_b/w_0)^2$, taking an infinite interaction $V\rightarrow\infty$. $g_R^{(2)}$ is already remarkably reduced from unity for $R_b\sim w_0$, and this ``anti-bunching'' becomes perfect ($g^{(2)}_R\rightarrow 0$) as $R_b$ increases further, indicating the impossibility to reflect two photons simultaneously. Intuitively, this effect arises because the detection of a first reflected photon implies that one atom within the illumination area had to be excited. However, the Rydberg blockade prohibits another atom from being simultaneously excited a radius $R_b$, which effectively punches a hole of that size in the mirror~(Fig.~\ref{Fig2}b). Note that without Rydberg interactions, detection of a reflected photon from an array of two-level atoms would only produce a single-atom hole in the mirror. This would result in a scaling $\g2R~\sim (1-1/N_i)^2$~(see Appendix~\ref{App:switch}), where $N_i\sim \pi \w0^2/d^2$ is approximately the number of atoms illuminated by the Gaussian beam~(inset of Fig.~\ref{Fig2}a). Here, $\g2R\sim 1$ implies that the first reflected photon has almost no impact on the ability of the mirror to reflect a second photon, \textit{i.e.} the mirror is highly linear.

These results suggest that a blockaded 2D array is the ``ultimate'' nonlinear element. In particular, it is lossless, unable to scatter light into modes beyond those sent in. Furthermore, the atom-light interaction heuristically is 100$\%$ efficient due to the combination of atom number and strong collective effects, as suggested by perfect reflection, but retains the nonlinearity of an ideal two-level system. While our analysis is phenomenological thus far, we now discuss how to realize the specific application of a high-fidelity single-photon switch.  \\   

\begin{figure}
    \centering
    \begin{minipage}{0.47\textwidth}
\centering
    \includegraphics[width=\textwidth]{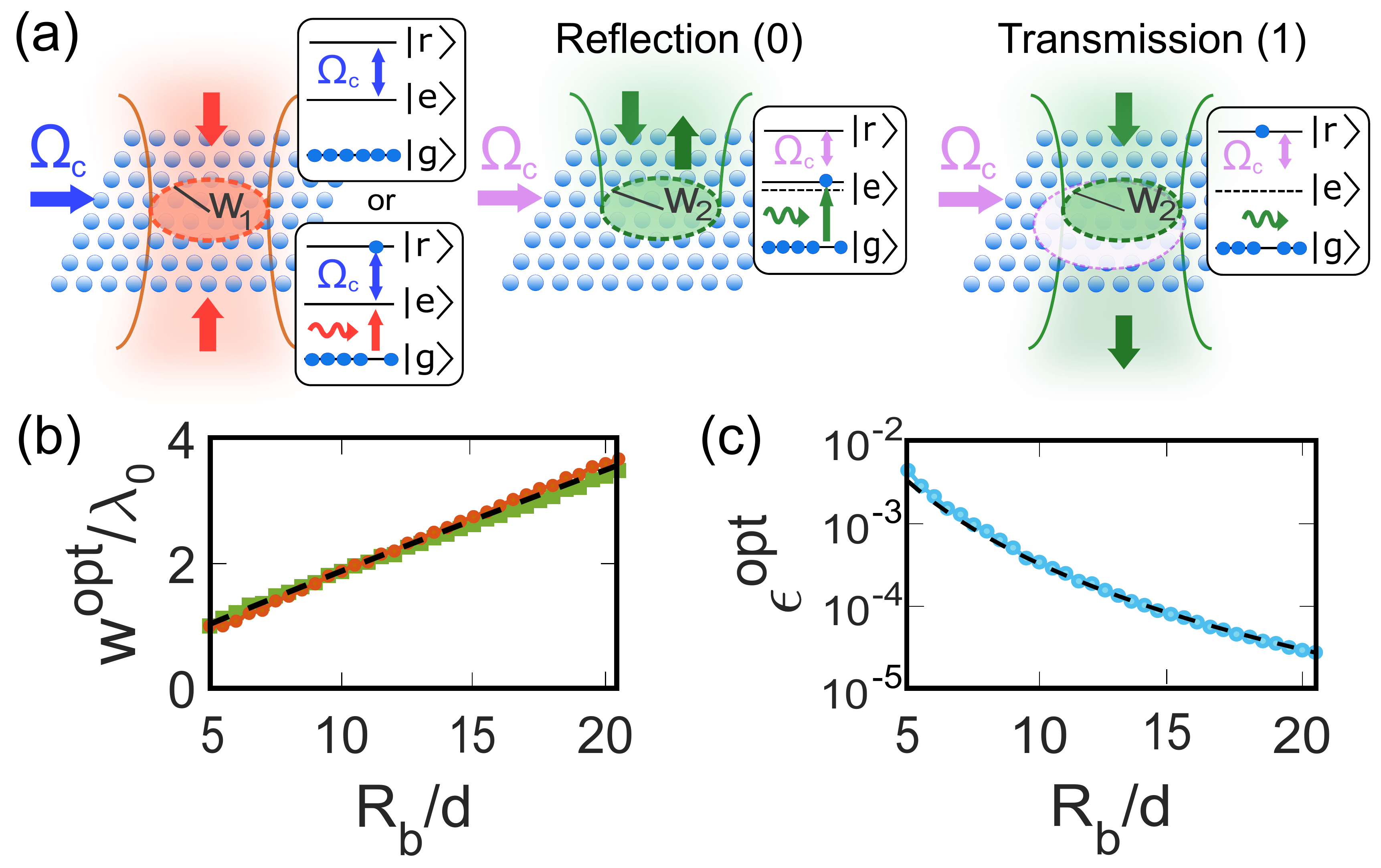}
    \end{minipage}
        \caption{ (a) Schematic of switch protocol. Initially (left panel), a gate pulse (red arrows) consisting of either zero (0) or one photon (1) is stored in the Rydberg level via a resonant control field (blue arrow). Afterwards, the control field is far detuned to induce Rydberg dressing ($|\delta_c|\gg|\Omega_c|$). A resonant signal photon (green arrows) is then sent from one direction. Depending on the gate photon number, the signal photon is reflected (0, middle panel) or transmitted (1, right). The transmission occurs as the stored Rydberg excitation punches a hole of radius $R_b$ in the array. Note that the Rydberg excitation (and thus hole center) is delocalized over a length $\sim w_1$ corresponding to the gate beam waist, as roughly illustrated by the off-center hole in transmission. (b) Optimal beam waists for the gate (red) and signal (green) photons that yield the minimal switch error. The dashed line corresponds to an analytical approximation, Eq. (\ref{eq:optimal_beam_waist}) in Appendix C. (c) Optimal switch error $\epsilon^{\rm opt}$ as a function of the blockade radius $R_b/d$. We also plot the analytical approximation Eq. (\ref{eq:gate_error}) in dashed lines. In the simulations, we consider a $N_a=41^2$ square array with lattice constant $d=\lambda_0/2$, and infinite Rydberg interaction strength.}
    \label{Fig3}
\end{figure}

\textit{Gate protocol.--} We first summarize the main steps of the single-photon switch, before analyzing them in detail. In this switch, the presence (absence) of a first ``gate'' photon conditions the array to be transmitting (reflecting) for a subsequent ``signal'' photon. Such a switch can be directly converted into a photon-photon gate with an additional classical beam-splitter, that converts the propagation direction into a conditional phase. We consider the three-level atom scheme from Fig. \ref{Fig1}a, where the gate and signal photons act on the $|g\rangle-|e\rangle$ transition (see Fig. \ref{Fig3}a). First, the gate pulse, which is assumed to consist of either zero or one photon, is split and sent toward the array from both directions, and then stored by applying a resonant control field ($\delta_c\approx-\delta$), creating either zero or one Rydberg excitation. Afterwards, the control field is detuned to achieve Rydberg dressing. The signal photon is finally sent from one direction with a frequency adjusted to compensate the Stark shift induced by the control field dressing. The signal then sees either a perfect resonant mirror or a large transmitting hole depending on the gate photon number. 

We now analyze each step separately. As in other works \cite{TNL17,ref:Gorshkov11}, we consider the limit to fidelity arising solely from the finite blockade radius, rather than limited atom number. We start with the storage of the gate pulse. Similar to the non-unity reflection in Eq. (\ref{eq:reflectance}), the finite beam waist $w_1$ of the gate photon results in a storage error of $1-\eta\approx C_s(d)\lambda_0^4/w_1^4$ \cite{MMA18}. As will be seen shortly, $w_1$ will become a non-trivial optimization parameter depending on the finite blockade radius $R_b$.

Without the gate photon, all atoms are in state $|g\rangle$ and the Rydberg dressing has no effect, such that an incoming signal photon with beam waist $w_2$ sees a mirror with an error in reflectance of $1-R \approx C_R(d)\lambda_0^4/w_2^4$ due to finite beam waist. In the case of a stored gate photon, the transmission error of the signal photon is somewhat more complex. In particular, storing the gate photon leads to a delocalized Rydberg excitation $|\Psi\rangle=\sum_jc_j\hat{\sigma}^{rg}_j|g\rangle$, where the state amplitudes follow the Gaussian profile of the gate field, $c_j\propto e^{-|\rb_j|^2/w_1^2}$. For a Rydberg dressing interaction $\hat{V}_{\rm Ryd}^{re}$ with infinite strength, $V\rightarrow\infty$, atoms within the blockade radius cannot be excited to state $|e\rangle$. A subsequent signal photon effectively sees a mirror with a hole of size $R_b$, but where the hole center ${\bf r}_j$ is in a delocalized superposition with weights $|c_j|^2$~(Fig.~\ref{Fig3}a). Note that the stored excitation itself introduces a single-atom hole in the mirror, but does not introduce any error as it belongs to the transmitting hole region. Furthermore, the signal photon response is linear optical, as the stored Rydberg excitation is static. It can be shown (see Appendix C) that the transmittance $T$ for the signal photon exactly corresponds to the weighted average 
 \begin{equation}
     T=\frac{\langle \hat{E}_{\rm T}^\dagger \hat{E}_{\rm T}\rangle_{\rm sc}}{\langle \hat{E}_{\rm T, in}^\dagger \hat{E}_{\rm T, in}\rangle_{\rm sc}}=\sum_i|c_i|^2\bar{T}(\rb_i,w_2,R_b),
     \label{eq:weighted_average}
 \end{equation}
where $\bar{T}(\rb_i,w_2,R_b)$ is the signal photon transmittance of an array with a hole of radius $R_b$ centered around atom $i$. Intuitively, the transmission will only be efficient if the uncertainty of where the hole is located, and the beam waist of the signal photon, are small, $w_{1,2}\lesssim R_b$.\\

We now quantify the overall fidelity of the switch, starting with a numerical optimization. Here, our only approximation involves the modeling of the Rydberg dressing interaction as $\hat{V}_{\rm Ryd}^{re} \approx  \sum_{i\neq j} V \Theta(R_b - r_{ij}) \hat{\sigma}_i^{rr} \hee_j$ with $V\rightarrow\infty$, while the storage efficiency $\eta$, and conditional reflectance and transmittance $R,T$ depending on the gate pulse are evaluated fully numerically. The total switch error $\epsilon$ is then the maximal error between storage/transmission and reflection, $\epsilon=\max (1-\eta T,1-R)$. In Figs.~\ref{Fig3}b,c we present the results for a 41x41 array (to avoid errors associated with finite array size) with lattice parameter $d=\lambda_0/2$. For different values of the blockade radius $R_b$, we plot the optimal beam waists $w_{1,2}^{\rm opt}$ in Fig.~\ref{Fig3}b, and the minimal error $\epsilon^{\text{opt}}$ in Fig.~\ref{Fig3}c. Separately, using a toy model based on the considerations above, we derive an analytical approximation of the error~(Appendix C),
 \begin{equation}
    \epsilon^{\text{opt}}(R_b,d)\approx C\frac{[1+\text{log}(R_b/d)]^2}{(R_b/d)^4},
    \label{eq:gate_error}
\end{equation}
which agrees well with the full numerics.  Notably, the $R_b^{-4}$ scaling significantly outperforms the best-known gate scaling $\propto R_b^{-3/2}$ in a disordered rEIT ensemble~\cite{TNL17}. In Appendix C, we show that this scaling can be realized in realistic settings, even when accounting for a finite Rydberg interaction strength and realistic potential shape.\\

\textit{Optical nonlinearities in the strong driving limit.--}
We now turn to the case of an arbitrarily large driving field, and study the optical response of the array depending on the driving power and blockade radius. In particular, we focus on the reflectance of light and on the photon loss, defined as the fraction of intensity scattered into non-detected modes beyond the reflected/transmitted Gaussian fields, i.e. $K=1-R-T$. We will show that this many-body quantum driven dissipative system can be understood in terms of a semi-classical model, based on holes being stochastically punched into the otherwise perfect mirror.
\begin{figure}[t]
\centering
\includegraphics[width=0.45\textwidth]{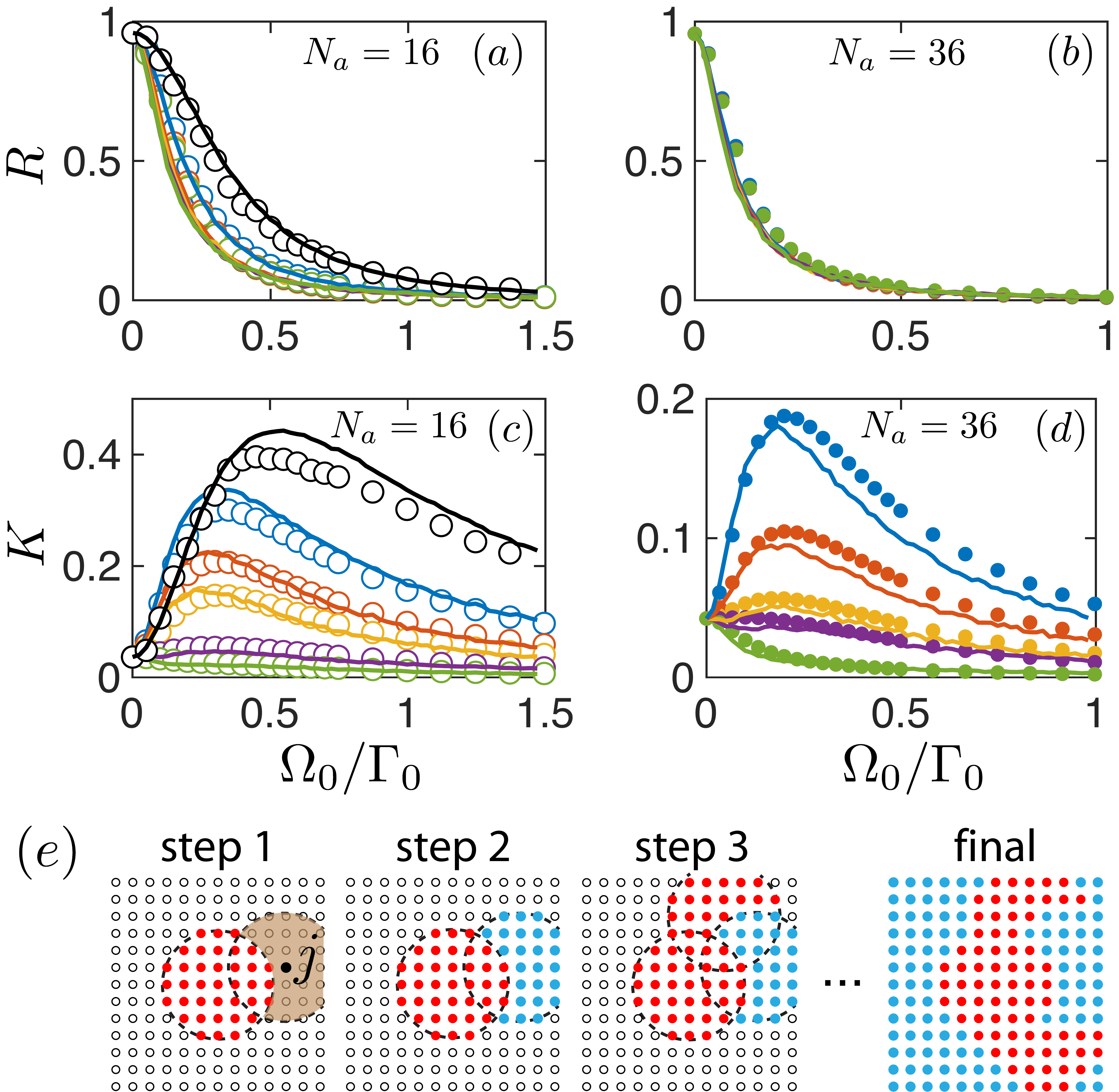}
\caption{(a)-(b) Reflectance and (c)-(d) photon loss versus Rabi frequency $\Omega_0/\Gamma_0$ for an incident Gaussian beam ($w_0 = 0.4 \sqrt{N_a} d$, $d = \lambda_0/2$) and an array with $N_a = 16$ or $N_a = 36$, as indicated. Different colors indicate different values of blockade radius $R_b$ (from top to bottom $R_b/d = 0, 1,\sqrt{2},2,\sqrt{5},3$ for $N_a=16$, and $R_b/d = \sqrt{8},\sqrt{10},4,\sqrt{18},5$ for $N_a=36$). Symbols denote the exact numerical calculation and solid lines are the semi-classical stochastic toy model. (e) Example of the steps to obtain a configuration used in the toy model. For each atom $j$ with no assigned state~(white circles), a blockade region~(brown shaded area in first panel) is defined, comprising all atoms within the blockade disk centered at $j$ but excluding those that have already been assigned a state in a previous step. An unassigned atom among this set is then randomly chosen, and the corresponding blockade region is probabilistically assigned to be reflecting (blue) or saturated (and thus effectively removed, red). The process is iterated until all atoms are assigned (right panel), and the optical properties of this configuration are calculated.}
\label{Fig4}
\end{figure} 

We first present numerical results for $N_a=16$ and $N_a=36$ square arrays. Numerically, we solve fully Eq.~\eqref{Eq:dynamics} for the steady-state density matrix, exploiting the infinite Rydberg blockade strength $V\rightarrow\infty$ to eliminate impossible-to-excite basis states and setting $\delta=\Delta_{{\bf k}_{\parallel}=0}$. In Figs.~\ref{Fig4}a,b, we see that the reflectance monotonically decreases with increasing $\Omega_0$ for all blockade radii, where $\Omega_0$ is the Rabi frequency at the center of the beam. This naturally arises from the saturation of the (Rydberg dressed super-)atoms. In contrast, the photon loss displays a non-monotonic behavior strongly depending on $R_b$ (Figs.~\ref{Fig4}c,d). Furthermore, the maximum loss $\Kmax$ occurs at some driving strength $|\Omega_0^{\rm max}|$, with both values depending non-trivially on $R_b$.

As previous intuition already suggests, it should be possible to approximately model this behavior in terms of transmitting/diffracting holes punched into a classical mirror (as we show in the final panel of Fig.~\ref{Fig4}e, where red and blue atoms illustrate effectively removed atoms and remaining mirror atoms, respectively). This assignment proceeds in a series of steps, starting with all atoms with no assigned state. Then, regions of the blockade radius size are randomly selected, and randomly assigned to be removed or kept depending on the probe field intensity in that region, which dictates their probability of saturation (see Appendix D). Once all atoms have been assigned~(Fig.~\ref{Fig4}e), we then calculate the corresponding classical (weak driving) loss and reflectance of this particular configuration, repeating and averaging over $\sim 5000$ configurations to obtain the loss and reflectance of the system.

Despite being a semi-classical stochastic model, it captures remarkably well the loss behavior for a wide range of cases. In Figs.~\ref{Fig4}a-d we overlay the previous numerical results with those predicted by the model (solid lines). If one further assumes that the system roughly consists of $N_d$ independent and non-overlapping blockade regions, which each have radius $R_b$ and see equal Rabi frequencies $\Omega$, and neglects corrections associated with finite array size, this model predicts that the maximum loss decreases as $\Kmax=(1-N_d^{-1})/2$ (see Appendix D). In particular, the loss vanishes when the system is fully blockaded and only a single excitation can be created ($N_d=1$). \\

\textit{Conclusions and outlook}
Here, we have shown that 2D atomic arrays with Rydberg interactions constitute a powerful platform for quantum nonlinear optics, and enable a gate with an error scaling better than that of a disordered atomic ensemble. Beyond that, this work also raises several interesting opportunities. First, this work should stimulate immediate possibilities for experiments, such as in quantum gas microscope setups, where efficient reflectance has already been shown~\cite{RWR20}, and where Rydberg dressing has separately already been implemented~\cite{ZvBS16,ZChR17}. Furthermore, taking into account previous results that arrays enable an exponentially better error scaling for quantum memories, we hypothesize that arrays could facilitate a better scaling for all major applications involving quantum atom-light interfaces, and it would be interesting to develop the optimized protocols for those. Finally, although we have focused on 2D arrays here, we anticipate that studying quantum nonlinear optics in other array geometries will generally be a rich area of future research. \\

\textit{Acknowledgements} 
The authors acknowledge stimulating discussions with I. Bloch, L. Tarruell, and F. Andreoli. DEC acknowledges support from the European Union's Horizon 2020 research and innovation programme, under European Research Council grant agreement No 639643 (FOQAL) and FET-Open grant agreement No 899275 (DAALI), AEI Europa Excelencia program (EUR2020-112155, project ENHANCE), MINECO Severo Ochoa program CEX2019-000910-S, Generalitat de Catalunya through the CERCA program, Fundaci{\'o} Privada Cellex, Fundaci{\'o} Mir-Puig, Plan Nacional Grant ALIQS (funded by MCIU, AEI, and FEDER), Fundaci{\'o}n Ram{\'o}n Areces Project CODEC, and Secretaria d'Universitats i Recerca del Departament d'Empresa i Coneixement de la Generalitat de Catalunya, co-funded by the European Union Regional Development Fund within the ERDF Operational Program of Catalunya (project QuantumCat, ref. 001-P-001644). DG aknowledges support from the Secretaria d’Universitats i Recerca de la Generalitat de Catalunya and the European Social Fund (2020
FI B 00196). MMC acknowledges funding from the European Union’s Horizon 2020 research and innovation program under the Marie Skłodowska-Curie Grant agreement No. 801110 and the Austrian Federal Ministry of Education, Science and Research (BMBWF). It reflects only the author’s view and the Agency is not responsible for any use that may be made of the information it contains. 

\appendix
\clearpage

\section{Rydberg dressing potential}
Here, we derive the $\hat{V}_{\rm Ryd}$ term in the Hamiltonian of Eq. (1b), which arises from coupling of the excited states $|e\rangle$ with a high-laying Rydberg level $|r\rangle$. We start from the Hamiltonian of the $|e\rangle-|r\rangle$ transition, which contains the Rydberg interaction and the control field terms
\begin{equation}
    \mathcal{H}^{er} = \sum_{i,j = 1, i<j}^{N_a}\frac{C_6}{r_{ij}^6}\hat{\sigma}_i^{rr}\hat{\sigma}_j^{rr}-\hbar\sum_{i=1}^{N_a}\left[\delta_\text{c}\hat{\sigma}_i^{rr}
    +\left(\Omega_c\hat{\sigma}_i^{re}+h.c.\right)\right],
    \label{eq:Rydberg_Hamiltonian}
\end{equation}
where $\delta_c$ is the control field detuning, $\Omega_c$ its Rabi frequency and $C_6$ a parameter related to the strength of the Rydberg interaction. Within the dressing regime ($|\delta_c|\gg \Omega_c$), the coupling term with $\Omega_c$ can be treated as a perturbation, such that the number of atoms in $|e\rangle$ and $|r\rangle$ are individually conserved, while the control field induces energy shifts within each number manifold. As discussed in the main text, we consider two different dressing scenarios. In the first one, no atoms are excited to the Rydberg state $|r\rangle$, so it is sufficient to consider the effect of the control field on the manifold of excited states. Using standard perturbation theory, the effective Hamiltonian for the $|e\rangle$ states up to fourth order is given by \cite{Macri-pohl,Macri-pohl2}
\begin{equation}
    \hat{V}_{\rm Ryd}^{ee}\approx\sum_{i=1}^{N_a}\hbar\Delta_{\rm ac}\hat{\sigma}_i^{ee}-\sum_{i,j = 1, i\not=j}^{N_a}\frac{\hbar|\Omega_c|^4}{\delta_c^3}\left(\frac{1}{1+\frac{r_{ij}^6}{R_b^6}} \right)\hat{\sigma}_i^{ee}\hat{\sigma}_j^{ee},
    \label{eq:correction1}
\end{equation}
where the blockade radius is defined as $R_b^6=C_6/2\hbar|\delta_c|$ with $\delta_c<0$ and $C_6>0$. From Eq. (\ref{eq:correction1}), we see that atoms far away ($r_{ij}\gg R_b$) experience the Stark shift $\Delta_{\rm ac}=|\Omega_c|^2/\delta_c-|\Omega_c|^4/\delta_c^3$ individually. On the other hand, the Rydberg interaction disrupts the dressing of the doubly excited states when two atoms in the $|e\rangle$ level are close to each other ($r_{ij}\ll R_b$). 

\begin{figure}[t]
\begin{minipage}{0.49\textwidth}
\centering
\includegraphics[width=\textwidth]{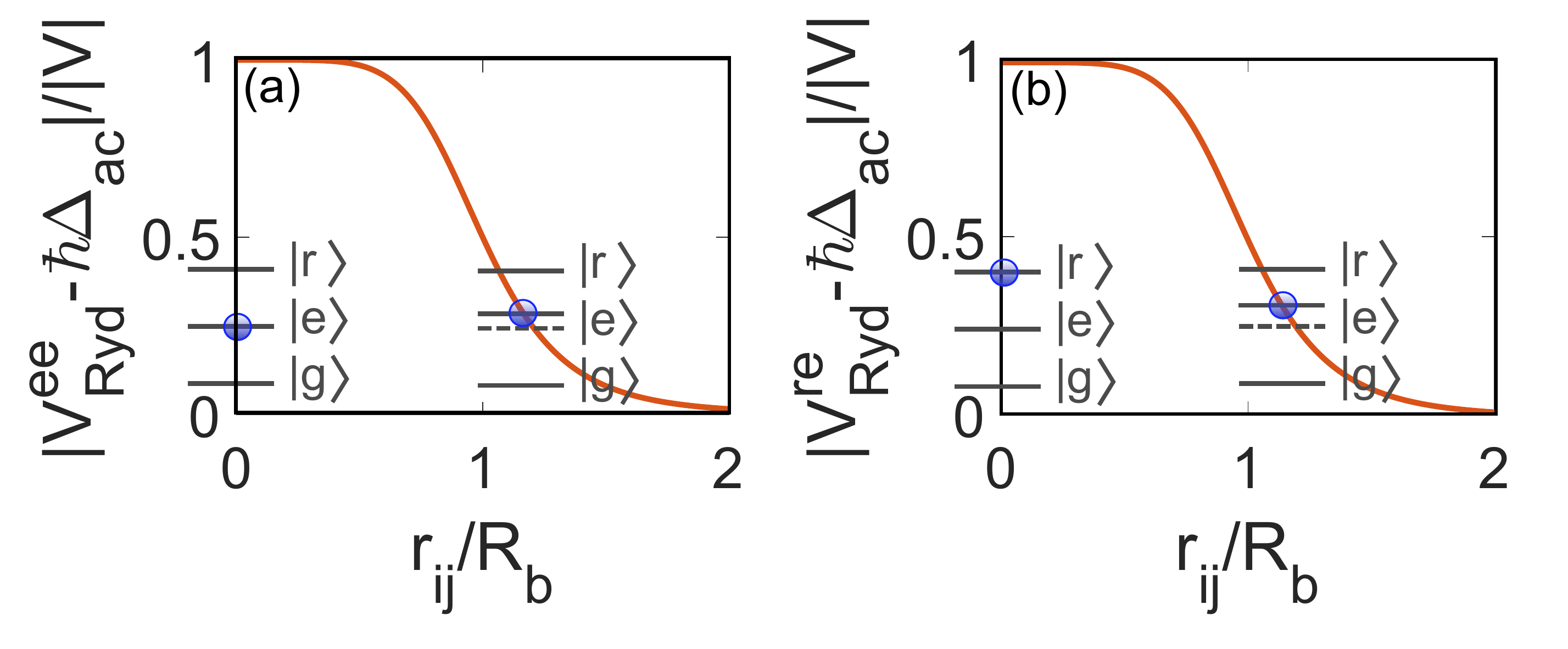}
\end{minipage}
\caption{(a) Rydberg dressing potential for a single atom in the excited state $|e\rangle$ in presence of another atom also in the state $|e\rangle$ at a distance $r_{ij}$ (schematically represented by the atomic levels $\{|g\rangle,|e\rangle,|r\rangle\}$ and the blue spheres). Specifically, we plot Eq. (\ref{eq:correction1}) after subtracting the single-atom Stark shift $\Delta_{\rm ac}$. (b) Same as before but now the atom is in presence of another atom in the Rydberg level. Thus, we plot Eq. (\ref{eq:correction2}). Both potentials have been rescaled respectively by $V=\hbar|\Omega_c|^4/\delta_c^3$ and $V=\hbar|\Omega_c|^2/\delta_c$. The distances have been rescaled by the blockade radius $R_b$. }
\label{Fig:Rydberg_dressing}
\end{figure}

In the second scenario, a single, immobile Rydberg excitation is first generated by storage of a photon. Then, other atoms can be excited to states $|e\rangle$ by a probe field. The effect of the stored Rydberg excitation on these excited states up to second order is described by the effective Hamiltonian 
\begin{equation}
    \hat{V}_{\rm Ryd}^{re}\approx\sum_{i,j = 1, i\not=j}^{N_a}\frac{\hbar|\Omega_c|^2}{\delta_c}\left(\frac{1}{1+2\frac{R_b^6}{r_{ij}^6}} \right)\hat{\sigma}_i^{rr}\hat{\sigma}_j^{ee}.
    \label{eq:correction2}
\end{equation}
 In particular, it can be seen that excited atoms that are much closer to the Rydberg excitation than a blockade radius do not experience a Stark shift at all. The potentials of the first and second dressing schemes from the main text are Eqs. (\ref{eq:correction1}) and (\ref{eq:correction2}) after incorporating the single-atom shift into the definition of the bare resonance frequency ($\omega_0\rightarrow \omega_0+\Delta_{ac}$), approximating the spatial dependence by the step function $\Theta(R_b^{\rm step}-r_{ij})$ \cite{Koscik2019} and defining $V=\hbar|\Omega_c|^4/\delta_c^3$ or $V=\hbar|\Omega_c|^2/\delta_c$, respectively. In Fig. \ref{Fig:Rydberg_dressing} we plot Eqs. (\ref{eq:correction1}) and (\ref{eq:correction2}) after subtracting the single-atom stark shift $\Delta_{\rm ac}$. 
 
We now discuss how to relate the blockade radius $R_b$ in the microscopic model to that in the step function approximation, ${R}_b^{\rm step}$. Physically, within the context of nonlinear optics, the step function represents the radius over which the array goes from transmitting to reflecting, due to the excited state shift imparted given either an excited or Rydberg atom at the center. From Eq.~(\ref{eq:r}), it can be seen that the excited state needs to be shifted by an amount $\sim\Gamma_{k_\parallel}$ for an array to change its response from being largely reflecting to transmitting. We thus define $|\hat{V}({R}_b^{\rm step})|= \Gamma_{k_\parallel=0}$, where $\hat{V}$ is either Eq. (\ref{eq:correction1}) or Eq. (\ref{eq:correction2}) for a single excited atom in presence of another atom in the state $|e\rangle$ or $|r\rangle$ (respectively), and after subtracting the single-atom stark shift $|\Delta_{\rm ac}|$. This leads to the relation
\begin{equation}
   \Tilde{R}_b^{\text{step}}\approx R_b\sqrt[6]{\kappa\left(\frac{V}{\Gamma_{\textbf{k}_\parallel=0}}-1\right)},
    \label{eq:rescaling_potential}
\end{equation}
where $V$ is that appropriate to the specific dressing scheme and $\kappa=1$ or $\kappa=2$ depending on using Eq. (\ref{eq:correction1}) or Eq. (\ref{eq:correction2}), respectively. In the main text, we simplify the discussion by treating $R_b^{\rm step}$ as an independent variable and dropping the \textit{step} label to reduce the complexity in notation. In Sec.~\ref{subsec:beyondstep}, we will see that the errors derived for the single-photon switch within the step function approximation agree well with those calculated from the actual microscopic potential.

\section{Steady state in the low driving intensity limit}
As stated in the main text, the dynamics of the atomic degrees of freedom are governed by the master equation Eq.~(\ref{Eq:dynamics}). An equivalent formulation of the master equation is the quantum jump formalism, wherein the system is described by a wave function $\ket{\Psi}$ that evolves  deterministically under the non-Hermitian Hamiltonian Eq.~(\ref{Eq:Heff}),
\begin{equation}
i \hbar \partial_t \ket{\Psi} = \Heff \ket{\Psi}. 
\label{Eq:dynamicsPsi}
\end{equation}
Within this formalism, the additional term Eq.(\ref{Eq:Jumps}) in the master equation is implemented via stochastic quantum jumps applied to the wave function. However, in the limit of weak input field, it is well-known~\cite{ref:textbook,MMA18} that the observables of interest (reflectance, transmittance and normalized two-photon correlation function) can be calculated neglecting the jumps. 

To be specific, we consider the dressing regime where no atoms are excited to $|r\rangle$, and the atoms can effectively be described as being two-level. We furthermore expand the wave function up to two atomic excitations, $\ket{\Psi} = c^{(g)} \ket{g} + \sum_{j=1}^{N_a} c_j^{(e)} \heg_j \ket{g} + \sum_{j,k =1, j<k}^{N_a} c_{jk}^{(2e)}  \heg_j \heg_k \ket{g}$. 

Applying Eq.~(\ref{Eq:dynamicsPsi}) to the wave function of the previous form leads to a linear system of differential equations for the coefficients: 
\begin{align}
i \dot{c}^{(g)} &= -\sum_{j=1}^{N_a} \Omega_j c^{(e)}_j, \notag\\
i \dot{c}_j^{(e)} &=-\Omega_j^* c^{(g)} -(\delta+i\Gamma_0/2) c^{(e)}_j +  \sum^{N_a}_{k=1,k\neq j} \mathcal{H}_{jk} c^{(e)}_k +\notag \\
& \qquad{} -\Omega_j\sum^{N_a}_{k=1, k\neq j} c_{jk}^{(2e)}\notag,\\
i \dot{c}_{jk}^{(2e)} &=-\left(\Omega^*_j c^{(e)}_k+\Omega^*_k c^{(e)}_j \right) -2(\delta+i\Gamma_0/2) c^{(2e)}_{jk} +\notag \\
& \qquad{} +\sum^{N_a}_{l=1,l\neq j} \mathcal{H}_{jl} c^{(2e)}_{lk}  + \sum^{N_a}_{l=1,l\neq k} \mathcal{H}_{lk} c^{(2e)}_{jl} +\cdots
\label{Eq:Dynamics_coef}
\end{align}
where the dots denote the contribution from the three excitations manifold, and where for simplicity we have defined $c^{(2e)}_{jk} \equiv c^{(2e)}_{kj}$ and the matrices $\mathcal{H}_{jk} = J^{jk}-i\Gamma^{jk}/2$. 

The steady state coefficients are found by imposing the time derivative to be zero in Eqs.~(\ref{Eq:Dynamics_coef}). Starting with all the atoms in the ground state ($\ket{\Psi(t=0)} = \ket{g}$), and replacing $c^{(g)} \approx 1$, it is possible to obtain an expression for the coefficients to lowest order in $\Omega_0/\Gamma_0$ ($c^{(e)}_j \sim O(\Omega_0/\Gamma_0)$ and $c^{(2e)}_{jk} \sim O(\Omega_0^2/\Gamma_0^2)$). 

Once the atomic degrees of freedom are solved, it is straightforward to reconstruct the light observables from the input-output relation Eq. (\ref{Eq:fields}). As a concrete example, we consider the reflectance, which is  defined as the rate of photons collected back into the same input mode, $R = \avg{\Er^\dagger \Er} / \avg{\Etin^\dagger \Etin}$. Here $\avg{\cdot}$ denotes the quantum mechanical expectation value on the atomic degrees of freedom, and thus the reflectance is directly given by the correlation functions $\avg{\heg_j \hge_k}$. 

\section{Single-photon switch}~\label{App:switch}
In this section, we provide a more detailed analysis of our proposed single-photon switch. First, we show that given the storage of a gate photon, the formula for the transmittance experienced by a subsequent signal photon indeed reduces to Eq.~(\ref{eq:weighted_average}) of the main text, involving a classical average over transmittance of an array with holes punched in different positions. We then derive the optimal switch error (Eq.~(\ref{eq:gate_error}) of the main text) by means of a toy model characterizing the signal photon transmission. Finally, we address the retrieval of the gate photon and discuss a realistic implementation of the switch by going beyond the step function approximation for the Rydberg dressing interaction, and considering the actual potential derived from perturbation theory, Eq.~(\ref{eq:correction2}).

\subsection{Formal theory of signal photon transmission}
Here we derive the transmittance $T$ from Eq. (\ref{eq:weighted_average}) in the main text. We begin by considering the state of the atomic array following storage of a gate photon, $|\Psi_0\rangle=\sum_{i}c_{i}^{(r)}\hat{\sigma}_i^{rg}|g\rangle$. Here, $c_i \propto e^{-|\rb_i|^2/w_1^2}$ follows the Gaussian spatial profile of the gate photon, and the wave function is normalized to unity in the case of perfect storage. Once the control field is detuned to create the Rydberg dressing, the excitation in $|r\rangle$ no longer evolves, and a signal photon is sent in the same Gaussian mode used for detection in transmission, $\hat{E}_{\rm T,in}$. Within a scattering formalism, this composite state $\hat{E}_{\rm T,in}^{\dagger}|\Psi_0,{\rm vac}\rangle$ formally transforms to
\begin{equation}
   |\Psi_{\text{sc}}\rangle=\sum_{i}\hat{\bf E}_{i}^{\dagger} c_i^{(r)}\hat{\sigma}_i^{rg}|g,{\rm vac}\rangle.
      \label{eq:state_after_scattering}
\end{equation}
Here, we have explicitly included the photonic component of the system wave function, and $\hat{\bf E}_{\rm i}$ is the mode into which the incoming photon scatters, if atom $i$ was in the Rydberg state. This mode could contain some non-zero projection into the detectable Gaussian reflection and transmission modes, as well as a continuum of other modes in $4\pi$~(thus representing loss). Note that in the limit of an infinite array, the spatial modes associated with $\hat{\bf E}_{i}$ for different atoms $i$ are identical, apart from a translation corresponding to the position of atom $i$ within the array.

To proceed further, we can formally decompose the mode $\hat{\bf E}_i$ into the detection mode in transmission, and orthogonal modes whose explicit forms are not needed,
\begin{equation}
   |\Psi_{\text{sc}}\rangle=\sum_{i}c_i^{(r)}\left(\sqrt{\bar{T}_i}e^{i\phi_i}\hat{\bf E}_{\rm T}^{\dagger}\hat{\sigma}_i^{rg}|g,{\rm vac}\rangle + |\rm Orthog.\rangle\right),
      \label{eq:state_after_scattering2}
\end{equation}
where we have expressed the overlap between $\hat{{\bf E}}_i$ and $\hat{{\bf E}}_{\rm T}$ in terms of a real number $\bar{T}(\rb_i,w_2,R_b)$ and phase $\phi_i(\rb_i,w_2,R_b)$. The overall transmittance of the signal photon into mode $\hat{{\bf E}}_{\rm T}$ is given by the total population in the state $\hat{{\bf E}}_{\rm T}^{\dagger}|{\rm vac}\rangle$, which reproduces Eq.~(\ref{eq:weighted_average}) in the main text. While this expression was formally derived by considering single-photon scattering, being a linear process, $\bar{T}(\rb_i,w_2,R_b)$ can also be calculated by considering the transmittance of weak coherent input light, which is what we implement numerically.

\subsection{Scattering properties of a single hole}
Physically, $\bar{T}(\rb_i,w_2,R_b)$ describes the transmittance of a signal photon of beam waist $w_2$, when atom ${\bf r}_i$ is in a Rydberg state and dressing interactions induce a blockade radius of size $R_b$. Here, we analyze more carefully the properties of scattering, in the simple limit where the interaction is approximated by a step function with infinite depth $V\rightarrow\infty$, so that no atoms within the blockade radius can be excited to state $|e\rangle$, thus punching a hole in the atomic array.

\begin{figure}[t]
\begin{minipage}{0.45\textwidth}
\centering
\includegraphics[width=\textwidth]{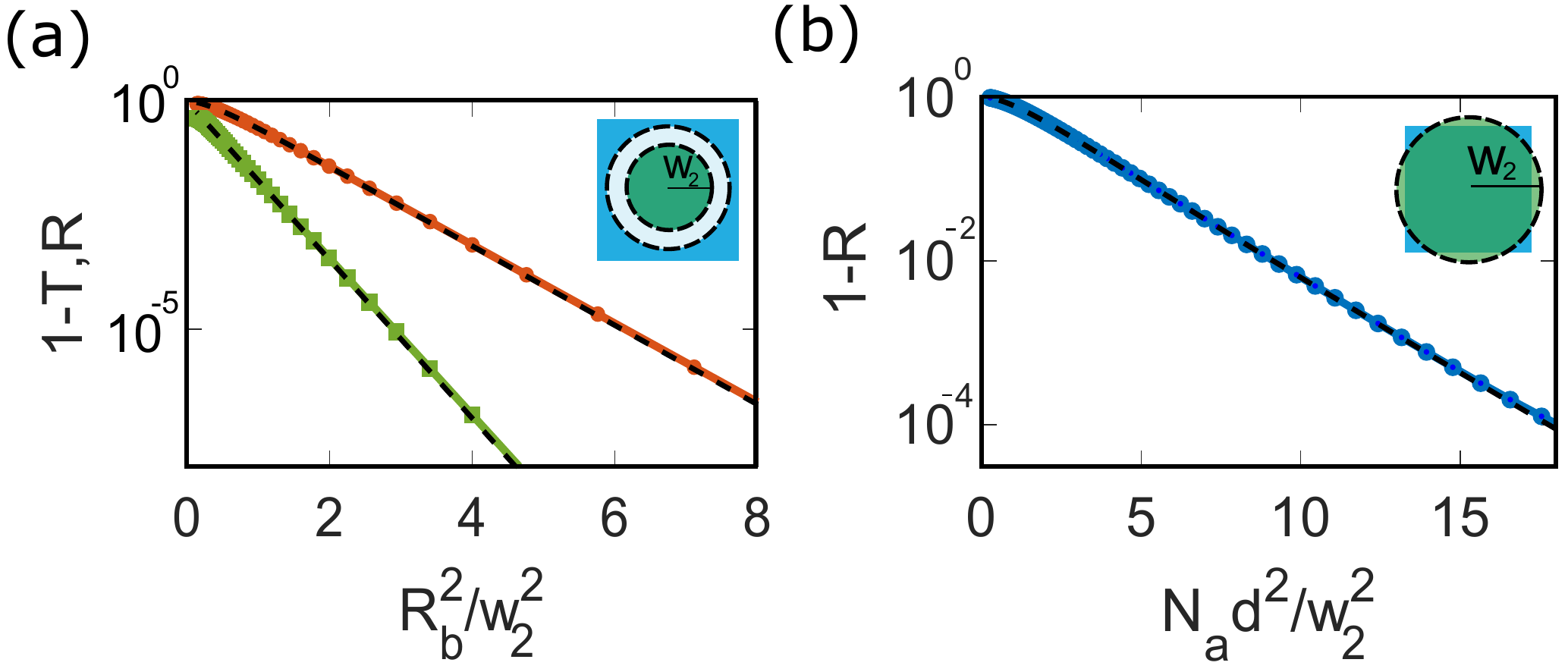}
\end{minipage}
\caption{(a) Numerically calculated classical transmittance (plotted as $1-T$, red circles) and reflectance $R$ (green squares) of a Gaussian field with beam waist $w_2$ illuminating a large~($N_a=41 \times 41$) array, with a hole~(atoms removed) of radius $R_b$ at its center. Here, $R,T$ are the projected reflectance and transmittance back into the Gaussian mode. We also plot the analytical formulas for $1-T$ using Eq. (\ref{eq:Transmittance_Centred_circular_aperture}), and $R=|e^{-2R_b^2/w_2^2}|^2$ in dashed black. (b) Reflectance~(plotted as $1-R$) of a Gaussian field resonantly driving a square atomic array of size $N_a=41 \times 41$, with varying beam waist $w_2$. The reflectance is back into the same Gaussian mode. We also plot Eq. (\ref{eq:reflectance}) from the main text (dashed lines). In both simulations, we consider a lattice constant $d=\lambda_0/2$.}
\label{Fig:Appendix_Transmission_Reflection}
\end{figure}

In the following, we approximate the array as a continuous mirror with a circular aperture, such that scattering can be treated in terms of classical diffraction theory. In particular, for an input Gaussian beam aligned with the aperture, it is well established that the fraction of power transmitted at the other side simply corresponds to the fraction of input power hitting the aperture, $P_{\rm tra}/P_{\rm in}=1-e^{-2R_b^2/w_2^2}$ \cite{Tanaka:85}. This can also be viewed as the overlap between the original Gaussian and transmitted modes. Note that this is the total transmitted power into \textit{all} modes, including both the original Gaussian and a set of orthogonal modes excited due to diffraction. In our case, however, we are only interested in the transmission back into the Gaussian mode. Due to reciprocity, this is given by the previous overlap squared, such that
\begin{equation}
    \bar{T}(\textbf{r}_i=0,R_b,w_2)=\left|1-e^{-2R_b^2/w_2^2}\right|^2,
    \label{eq:Transmittance_Centred_circular_aperture}
\end{equation} 
where we have neglected corrections arising from the finite total array size. Similarly, the fraction of power reflected by the mirror with the aperture into all modes is $P_{\rm rfl}/P_{\rm in}=e^{-2R_b^2/w_2^2}$, while projecting back into the Gaussian mode gives $R\approx |e^{-2R_b^2/w_2^2}|^2$. In Fig. \ref{Fig:Appendix_Transmission_Reflection}a, we perform a full numerical simulation of an array of $N_a=41^2$ atoms illuminated with a weak, resonant Gaussian beam of waist $w_2=5d$, and with all atoms within a radius $R_b$ of the origin removed. We see that the numerically evaluated reflectance and transmittance as a function of radius $R_b$ agree well with the approximate formulas for $\bar{T}_i$ and $R$ derived above.\\

These results can readily be generalized by replacing $e^{-2R_b^2/w_2^2}\rightarrow \nu$ in the previous formulas, where $1-\nu$ physically describes the fraction of input beam power hitting a hole in an array of any position and size. As one relevant consequence, to be used later, the loss, defined as the fraction of power that is neither transmitted nor reflected into the original Gaussian mode, is given by $K=1-T-R=2\nu (1-\nu)$.

For completeness, we use the previous arguments to justify the reflectance in Eq. (\ref{eq:reflectance}). Considering a finite square mirror of size $(Nd)^2$ illuminated by a Gaussian beam, the fraction of power reflected is $P_{\rm rfl}/P_{\rm in}=\text{ erf}^2(Nd/\sqrt{2}w_2)$, where the rest of the light is leaked at the edges of the mirror. The reflectance back into the Gaussian mode is then this quantity squared, corresponding to the first term in Eq. (\ref{eq:reflectance}). In Fig.~\ref{Fig:Appendix_Transmission_Reflection}b, we see good agreement between this analytical result and the numerical evaluation of the reflectance from a finite array.
\subsection{Approximate model for signal photon transmission}
\begin{figure}[t]
\begin{minipage}{0.48\textwidth}
\centering
\includegraphics[width=\textwidth]{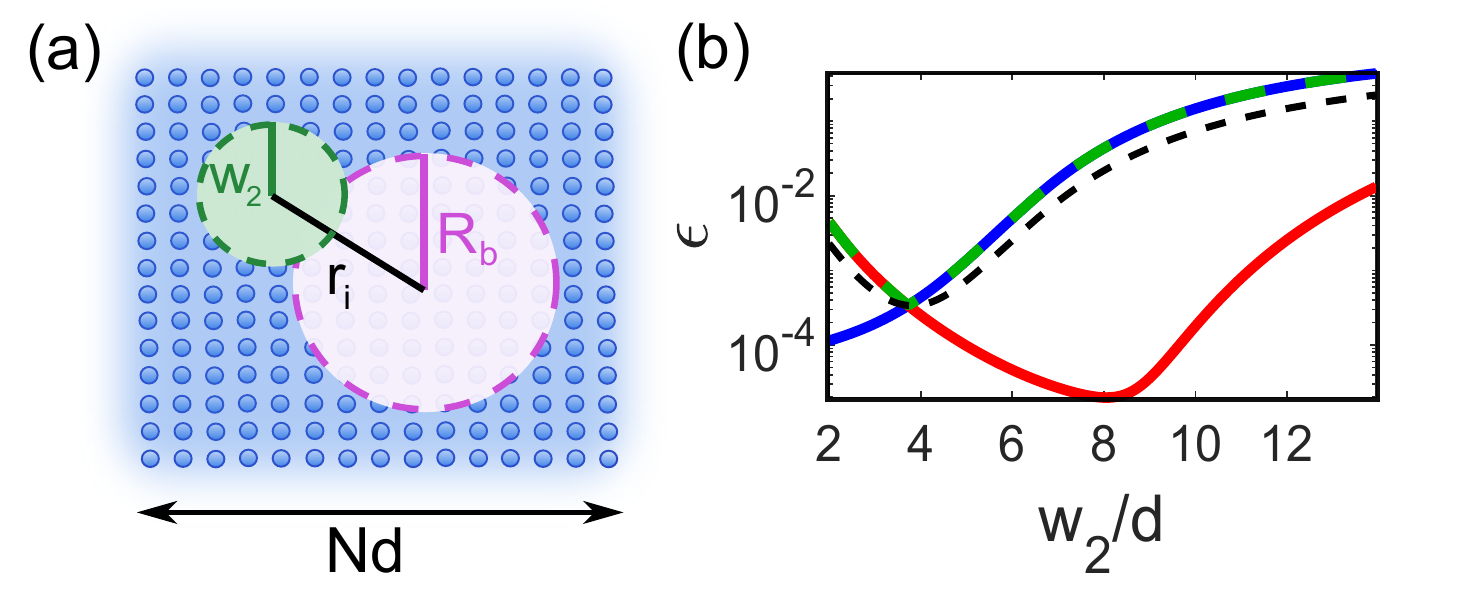}
\end{minipage}
\caption{(a) Sketch of the toy model used to calculate signal photon transmittance. We assume that the signal photon consists of a top-hat profile with radius $w_2$. The transmittance then reduces to finding the intersection area between this circle of radius $w_2$~(green), and the hole created by a locally stored Rydberg excitation (pink circle of radius $R_b$), which are separated by a distance $|\rb_i|$. (b) Reflection error $1-R$ (red) and storage/transmission error $1-\eta T$ (blue) of the switch as a function of the signal photon beam waist $w_2$ and after optimizing over the gate photon beam waist $w_1$.  The total switch error $\epsilon$ is defined as the maximal error between transmission and reflection (dashed green curve). We also plot the function $f(w)=(\epsilon_T+\epsilon_R)/2$ (dashed black), whose minimum coincides with the optimal (smallest) switch error. The specific values are obtained from a numerical simulation with $R_b=10d$, $N_a=41^2$ and $d=\lambda_0/2$.} 
\label{Fig:Geometric_toy_model}
\end{figure}

Unfortunately, while $\bar{T}(\rb_i=0)$ through a single hole aligned with a Gaussian beam admits a simple, closed-form expression, we do not find a simple solution for $\nu$ once the hole is misaligned. Furthermore, the signal photon transmittance of Eq.~(\ref{eq:weighted_average}) involves a weighted sum of transmittance through off-center holes. We thus consider a toy model that captures the essential physics, in order to derive an approximate scaling. In particular, we assume the signal photon has a top-hat spatial profile, with radius $w_2$. Then, the problem becomes purely geometrical, as the transmittance involves the overlap area between two circles of radius $R_b$ and $w_2$, separated by a distance $|\rb_i|$ (Fig. \ref{Fig:Geometric_toy_model}a). Assuming that $R_b>w_2$, we identify three regimes of interest (Fig. \ref{Fig:Geometric_toy_model}b):
\begin{equation}
     \bar{T}(\rb_i,w_2,R_b)\approx\begin{cases}
               1\qquad \text{for $R_b-w_2>|r_i|$},\\
               0\qquad \text{for $|\rb_i|>R_b+w_2$},\\
               I_B(\rb_i,w_2,R_b)\qquad \text{else},\\
            \end{cases}
            \label{eq:Ansatz_transmission}
\end{equation}
where $I_B\in(0,1)$ is a polynomial whose explicit form will not be relevant here. Substituting the ansatz from Eq. (\ref{eq:Ansatz_transmission}) into Eq. (\ref{eq:weighted_average}), one finds 

\begin{equation}
   T=\int_0^{R_b-w_2} 2\pi \rho |c|^2\ \text{d}\rho+\int_{R_b-w_2}^{R_b+w_2} 2\pi \rho I_B|c|^2\ \text{d}\rho,
   \label{eq:integral}
\end{equation}
where we have taken the continuous limit, and $|c|^2=2 e^{-2\rho^2/w_1^2}/(\pi w_1^2)$ after normalizing the Rydberg population to unity. Finally, considering $R_b\gg w_2$ to neglect the second integral in Eq. (\ref{eq:integral}), one obtains
\begin{equation}
    T(w_1,w_2,R_b)\approx 1-e^{-2(R_b-w_2)^2/w_1^2}.
    \label{eq:T_simplified_toy_model}
\end{equation}
Next, we combine the previous Eq. (\ref{eq:T_simplified_toy_model}) with the expressions for $\eta$ and $R$ discussed in the main text to find an analytical approximation for the optimal switch error, which was defined as the maximal error between storage/transmission ($\epsilon_t=1-\eta T$) and reflection ($\epsilon_r=1-R$). For simplicity, we will use the ansatz $w_1=w_2=w$, which is motivated by the results from Fig. \ref{Fig3}c in the main text. Intuitively, one expects the optimal switch to have the same error in transmission and reflection. This is illustrated in Fig. \ref{Fig:Geometric_toy_model}b for the particular case of $R_b=10d$, where we plot the numerically calculated reflection~(red) and storage/transmission error~(blue) for different beam waists $w_2$, with the minimum error occurring at the intersection of these curves. However, solving $\epsilon_r(w^{opt})=\epsilon_t(w^{opt})$ leads to a transcendental equation for $w^{opt}$. To circumvent this, we can instead minimize the function $f(w)=(\epsilon_r+\epsilon_t)/2$, whose minimum is also at $w^{opt}$ (see Fig. \ref{Fig:Geometric_toy_model}b). Considering only the leading terms, the optimal beam waist is then given by
\begin{equation}
    w^{\rm opt}(R_b,d)\approx\frac{R_b}{1+\sqrt{\log(C_\epsilon R_b/d)}},
    \label{eq:optimal_beam_waist}
\end{equation}
where the constant $C_\epsilon(d)$ can be obtained by fitting the data. For $d=\lambda_0/2$, we get $C_\epsilon\approx2$. The optimal switch error in Eq. (\ref{eq:gate_error}) can be obtained by substituting $w^{\rm opt}$ from Eq. (\ref{eq:optimal_beam_waist}) into $\epsilon_R(w)\approx C_R(d)\lambda_0^4/w^4$. In the main text, we approximate the resulting expression to have a single constant $C(d)$, instead of two parameters $C_R(d)$ and $C_\epsilon(d)$ that unnecessarily complicate the discussion.
\begin{figure}[t]
\begin{minipage}{0.48\textwidth}
\centering
\includegraphics[width=0.60\textwidth]{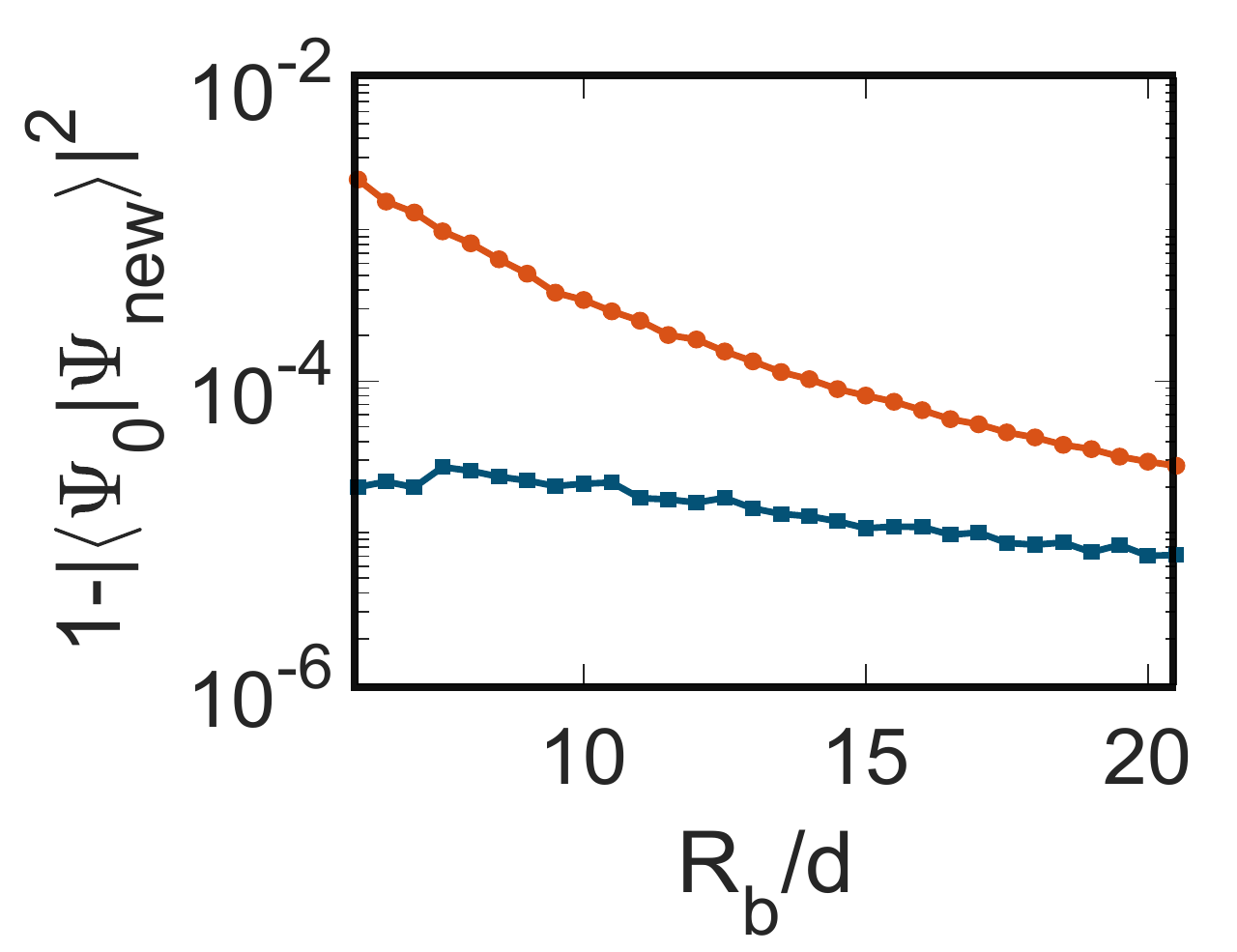}
\end{minipage}
\caption{Overlap error (blue squares) between the state of the array following storage $|\Psi_0\rangle$ and the state $|\Psi_{\rm new}\rangle$ after the detection of a signal photon in transmission. We also plot the gate error (orange circles) from Fig. \ref{Fig3} for comparison. In the simulations, we consider a $N_a=41^2$ square array of lattice constant $d=\lambda_0/2$. The beam waists are the ones that optimize the switch error, which allows us to plot the data as a function of the blockade radius.}
\label{Fig_Appendix_retrieval_error}
\end{figure}
\subsection{Retrieval of the gate photon}
To operate our single-photon switch as a quantum gate, one must also consider the error associated with retrieving the first gate photon after the signal photon has been scattered. In particular, while storing and directly retrieving the gate photon (absent the signal photon) would have the same efficiency $\eta$~\cite{MMA18}, from Eq.~(\ref{eq:state_after_scattering2}), one sees that conditioned on the signal photon being transmitted into the desired mode $\hat{\bf E}_T$, the remaining Rydberg spin wave is altered, from $|\Psi_0\rangle = \sum_i c_i^{(r)} \hat{\sigma}_i^{rg} |g\rangle$ to $|\Psi_{\rm new}\rangle = \mathcal{N} \sum_i c_i^{(r)} \sqrt{\bar{T}_i} e^{i\phi_i} \hat{\sigma}_i^{rg} |g\rangle$. Here, $\mathcal{N}$ is a normalization factor such that $|\Psi_{\rm new}\rangle$ has unit norm (the smaller than unity norm without this factor describes the probability that the signal photon was not successfully transmitted, and its error has already been included in previous analysis). We now show that the additional error in retrieval due to this change of state is negligible compared to the total switch error previously calculated.

From time reversal symmetry, the excitation in the state $|\Psi_0\rangle$ can be retrieved with efficiency $\eta$, which we know is optimal. Then, any other state $|\Psi_\perp^i\rangle$ orthogonal to $|\Psi_0\rangle$ will have a lower retrieval efficiency $\eta_\perp^i$. Decomposing $|\Psi_{\text{new}}\rangle$ in the basis of $|\Psi_0\rangle$ plus other orthogonal states, one can express the retrieval efficiency of the distorted excitation as $\eta_{\text{new}}=\eta|\langle \Psi_0|\Psi_{\text{new}}\rangle|^2+\sum_i\eta_{\perp}^i|\langle \Psi_\perp^i|\Psi_{\text{new}}\rangle|^2$. In Fig. \ref{Fig_Appendix_retrieval_error}, we show that the distortion in the stored excitation (quantified by the overlap error $1-|\langle \Psi_0|\Psi_{\text{new}}\rangle|^2$) is smaller than the optimal gate error. Thus, even in the worst case scenario where $\eta_\perp^i=0$, one can still approximate $\eta_{\rm new}\approx \eta$.

\subsection{Switch error beyond the step-function approximation}\label{subsec:beyondstep}

So far, we have approximated the Rydberg dressing interaction as a step-function potential with infinite depth $V\rightarrow\infty$. In the following, we will show that the optimal switch error from Eq. (\ref{eq:gate_error}) can be achieved under realistic conditions, i.e. taking into account the real potential derived from perturbation theory, Eq. (\ref{eq:correction2}), and for a finite interaction strength $V=|\Omega_c|^2/\delta_c$. 

\begin{figure}[t]
\begin{minipage}{0.42\textwidth}
\centering
\includegraphics[width=\textwidth]{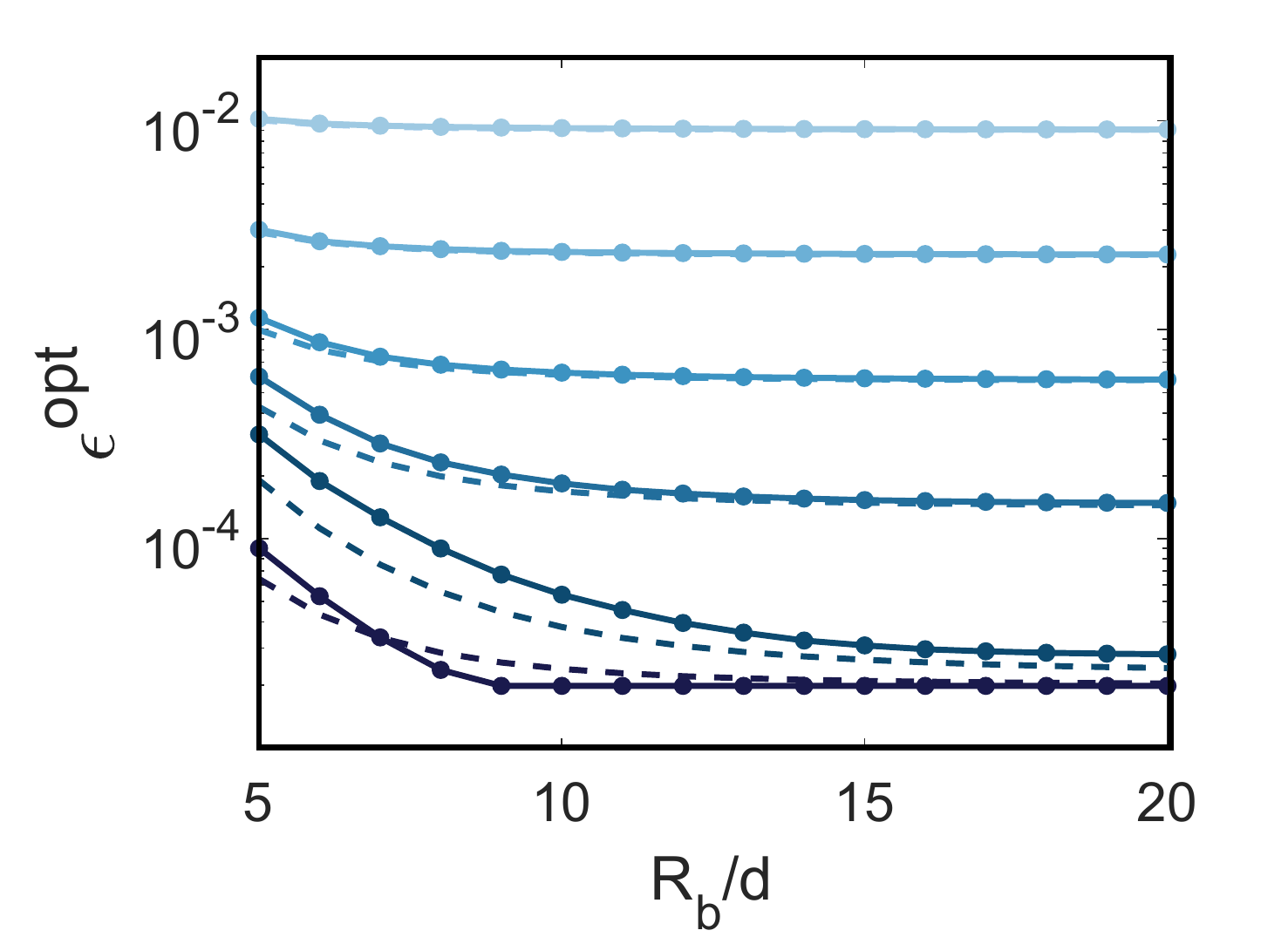}
\end{minipage}
\caption{Optimal gate error as a function of the microscopic blockade radius $R_b$ considering the real potential derived from perturbation theory [Eq. (\ref{eq:correction2})]. Each color represents a different interaction strength $V=[5,10,20,40,10^2,10^3]\Gamma_0$, respectively from light to dark blue. The solid lines are the results of a numerical simulation with a square array with $N_a=41^2$ and lattice parameter $d=\lambda_0/2$. The dashed lines are obtained substituting Eq. (\ref{eq:rescaling_potential}) into Eq. (\ref{eq:gate_error}) from the main text and adding the errors associated to finite $V$ and finite array size.}
\label{Fig:Appendix_Switch_real_potential}
\end{figure}

First, we numerically optimize the switch error considering the real potential (i.e. without the step function approximation). In Fig. \ref{Fig:Appendix_Switch_real_potential} we show the resulting $\epsilon^{\rm opt}$~(solid lines) as a function of the microscopic blockade radius $R_b$ from Eq. (\ref{eq:correction2}), and for different $V$. For now, we take $R_b$ and $V$ to be independent parameters, while their dependence on laser parameters and principal quantum number is discussed later. Interestingly, now $\epsilon^{\rm opt}$ is not arbitrarily small for increasing blockade radius as it saturates to a specific value in the limit $R_b\rightarrow\infty$. For moderate interaction strengths ($V\lesssim100\ \Gamma_0$), the saturation arises from the non-zero reflectance of the atoms within the blockaded region, which are not completely shifted out of resonance due to $V$ being finite. According to Eq. (\ref{eq:r}), this introduces an error in transmission given by $\epsilon_V\sim1/(1+4|V|^2/|\Gamma_{k_\parallel=0}|^2)$ that lower bounds $\epsilon^{\rm opt}$ as a function of $V$. On the other hand, for very large interaction strengths ($V\gtrsim10^3\Gamma_0$), $\epsilon_V$ becomes negligible and transmission can be considered as perfect. However, the saturation still appears, now due to the finite array size error $\epsilon_N$ that prevents perfect reflection, as we discussed in the main text. Thus, according to Eq. (\ref{eq:reflectance}), even if $V\rightarrow\infty$ and $R_b\rightarrow\infty$, the switch error will still be fundamentally limited by $\epsilon_N=1-\text{erf}^4(Nd/\sqrt{2}w_2)$.

To generalize Eq. (\ref{eq:gate_error}) beyond the step function approximation and to validate our interpretation of the solid lines in Fig. \ref{Fig:Appendix_Switch_real_potential}, we add the aforementioned errors $\epsilon_V$ and $\epsilon_N$ to the optimal switch error $\epsilon(R_b^{\rm step})$ from the main text. In addition, we use Eq. (\ref{eq:rescaling_potential}) to express $R_b^{\rm step}$ as a function of the microscopic $R_b$, in order to compare on the same plot. In Fig. \ref{Fig:Appendix_Switch_real_potential} we plot $\epsilon^{\rm opt}(R_b^{\rm step})+\epsilon_V+\epsilon_N$~(dashed lines), in terms of the microscopic $R_b$. Overall we observe a good agreement between solid and dashed lines, which validates our previous claims.\\

\begin{figure}[t]

\begin{minipage}{0.49\textwidth}
\centering
\includegraphics[width=\textwidth]{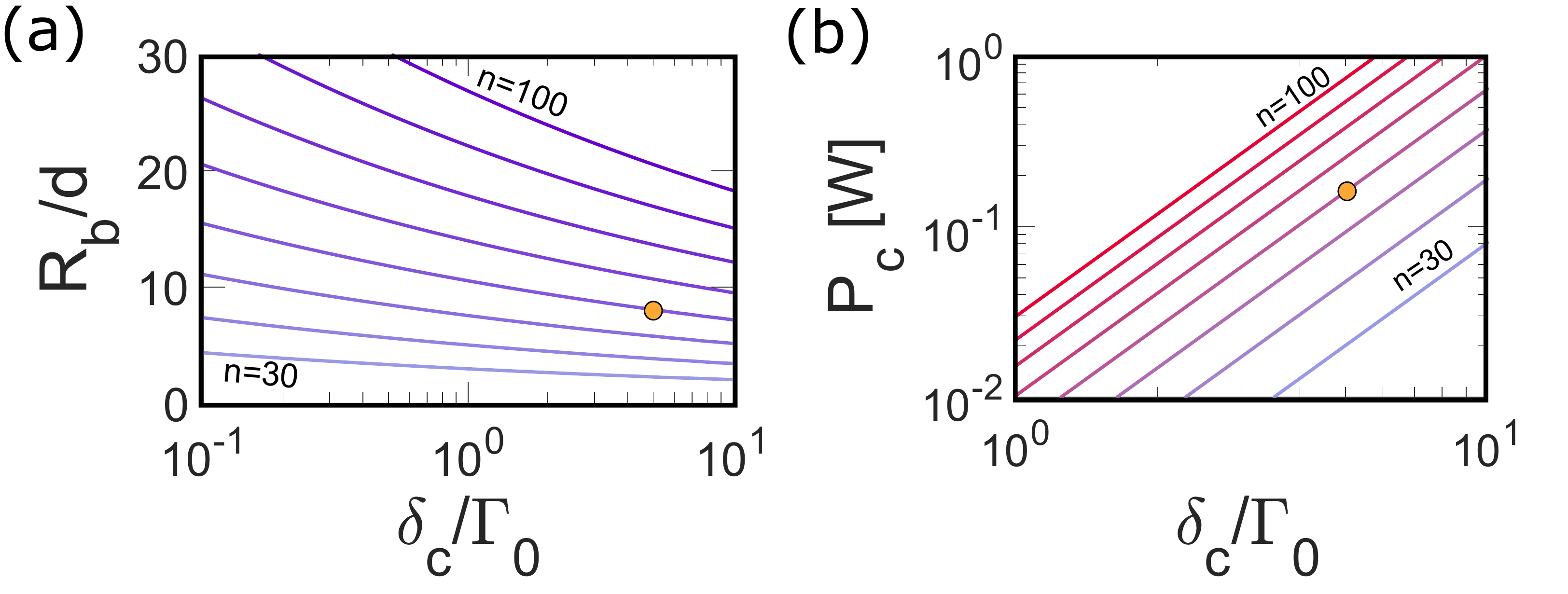}
\end{minipage}
\caption{Realistic values for the blockade radius $R_b$ (a) and control field power $P_c$ (b) as a function of the control field detuning $\delta_c$, rescaled by the linewidth $\Gamma_0=2\pi \times 6.065$ MHz of the $|e\rangle\rightarrow|g\rangle$ transition of ${}^{87}$Rb. To obtain these plots, we work at the limit of validity of perturbation theory, where $\Omega_c\approx\delta_c$ and $V\approx \delta_c$. Each color represents a different Rydberg state with principal quantum number $n\in\{30,40,...,100\}$, respectively from light grey to either purple or red. The yellow circle indicates the case of a photon switch with $99\%$ efficiency.}
\label{Fig_Appendix_realistic_values}
\vspace{-1em}
\end{figure}
Finally, we discuss the experimental feasibility of the values for $R_b$ and $V$ used in Fig. \ref{Fig:Appendix_Switch_real_potential}. As a concrete example, we will consider the use of $^{87}\text{Rb}$ atoms and assume the Rydberg states are reached following the widely used transition scheme $|g\rangle=|5S_{1/2}\rangle \rightarrow |e\rangle=|5P_{3/2}\rangle\rightarrow |r\rangle=|nS_{1/2}\rangle$~\cite{ref:Peyronel2012,ref:Distante2017,ref:Ryabtsev_2016}, with the ground-excited state transition wavelength of $\lambda_0=780.0$ nm, lattice constant $d=\lambda_0/2$, and excited state spontaneous emission rate $\Gamma_0=2\pi\times 6.065$ MHz~\cite{ref:Steck}. More specifically, an approximate two-level transition can be realized by utilizing a cycling transition, where the ground and excited states have maximum angular momentum and are connected via a circularly polarized transition, $|g\rangle= |F=2, m_F=2\rangle$ and $|e\rangle = |F=3,m_F=3\rangle$. It is known that dipole-dipole interactions in the presence of hyperfine structure can induce transitions out of this manifold, as the re-scattered field seen by a ground-state atom, coming from another excited atom, does not necessarily have the same circular polarization \cite{ref:AnaPNAS}. However, the probability of this is suppressed both by the squared ratio of Clebsch-Gordan coefficients between the undesired and desired transitions~($1/15$ for ${}^{87}$Rb), and by the application of magnetic fields, to realize a differential Zeeman shift $\delta_{{\rm Zeeman}}$ between the undesired and desired transition resonance frequencies~(with a corresponding suppression $\sim (\Gamma_0/\delta_{{\rm Zeeman}})^2$).

Within such a scheme, we now consider the Rydberg interaction properties. To reduce the number of parameters, we will work in the limit of validity of perturbation theory for the Rydberg dressing scheme, where $\Omega_c \approx \delta_c$ and $V\approx \delta_c$. In Fig. \ref{Fig_Appendix_realistic_values}a we plot the blockade radius $R_b$ as a function of the detuning $\delta_c$ for different principal quantum numbers $n$. The values for the $C_6$ coefficient have been obtained by fitting the data from Refs. \cite{ref:32,ref:Browaeys_2016,ref:Ds} with the function $C_6(n)=C_0n^{11}$. Additionally, we also calculate the control field power $P_c=2\epsilon_0c|\delta_c|^2\hbar^2A/|\db_{er}|^2$ required to achieve a specific $V\sim\delta_c$, where the surface illuminated by the control field $A$ is taken to cover the array in the simulations $A=\pi(20d)^2$. The value for the dipole moment $\db_{er} =(43/n)^3\cdot 0.0103\text{a}_0e$ is taken from Ref. \cite{ref:experimental_data_Rydberg} and rescaled to arbitrary $n$. Combining the results from Fig. \ref{Fig:Appendix_Switch_real_potential} and \ref{Fig_Appendix_realistic_values}, we conclude that a single-photon switch with $99\%$ efficiency can be already achieved with a control field of $\delta_c\sim 2\pi \times 30 $MHz, $P_c\sim 100$mW and $n\sim60$, which are consistent with the typical values currently used in state-of-the-art experiments \cite{ref:Browaeys_2016,ref:BlochExperiment}.

\section{Strong driving limit}
\subsection{Stochastic semi-classical model}
Here, we describe in more detail the stochastic semi-classical model that we have used to model the reflectance and loss of a 2D array in the strong driving limit. In the following we specify the procedure that determines how atoms are assigned to be saturated and effectively removed or not (as illustrated in the piece-wise steps of Fig.~\ref{Fig4}e, where red and blue atoms illustrate removed atoms and remaining mirror atoms, respectively).

This assignment proceeds in a series of steps, starting with all atoms with no assigned state. In each step, we apply the following rules: (i) we take all atoms $j$ that have not yet been assigned a state, and define the blockade region containing $N_b^j$ atoms as the intersection between the subset of atoms with no assigned state and the blockade radius centered at $j$ (see as an example the brown enclosed region in the first panel of Fig.~\ref{Fig4}e). (ii) A single atom $j$ from this subset is randomly chosen following the probability $\Omega_j^2 N_b^j /\sum_j \Omega_j^2 N_b^j$. Here $\Omega_j$ is the local Rabi frequency at $\rb_j$, and this quantity accounts for the likelihood that a dressed Rydberg superatom centered around $j$ becomes excited. (iii) The $N_b^j$ atoms contained in the corresponding blockade region of $j$ are randomly assigned to be saturated or unsaturated according to the probability $p_j= s_j/(1+s_j)$. The saturation parameter coincides with the usual one for a single two-level system, but with a collectively enhanced Rabi frequency $\sqrt{N_b^j}\Omega_j$ and collectively modified decay rates and resonance shifts, $s_j = 8 N_b^j \Omega_j^2/[\Gamma_{\kpar=0}^2+4(\delta -\Delta_{\kpar=0})^2]$. Once all atoms have been assigned, we then numerically calculate the corresponding linear classical loss and reflectance of this particular configuration in the weak driving limit, by effectively removing all the atoms in the saturated regions. We then repeat, sampling over $\sim 5000$ configurations and averaging to obtain the loss and reflectance of the system, respectively.

\subsection{Approximate analytical behavior of semi-classical model}
\label{AppC2}
While the reflectance and loss of the previously discussed toy model are still calculated numerically, it is possible to obtain analytical approximations with a few additional assumptions. In particular, we assume that the system is illuminated by a beam of area $A$, and can thus roughly be divided into $N_d = A/N_b d^2$ independent and non-overlapping blockade regions, each of radius $R_b$ and $N_b\sim \pi R_b^2/d^2$ atoms, which all see equal Rabi frequency $\Omega$ and individually saturate with probability $p=s/(1+s)$, where $s = 8N_b \Omega^2 /[\Gamma_{\kpar=0}^2+4(\delta-\Delta_{\kpar=0})^2]$ is the saturation parameter for the uniform field defined analogously as before.

The probability for $m$ of these regions to be saturated and transmit light then follows a binomial distribution, $\mathcal{P} (m)= {\binom {N_d} m }~p^m (1-p)^{N_d - m}$. Each of these configurations diffracts light as if it was a classical mirror with $m$ punched holes. We then generalize the result for the loss caused by a single hole in Appendix C.2 to the case of multiple holes; in particular, with a total fractional area $\nu=m/N_d$ removed from the mirror, the loss follows $K_{\rm cl} \approx 2 \nu (1-\nu)$, largely independently of the size and number of the regions. The total loss is then evaluated as a statistical average over any possible number of holes, $K = \sum_{m=0}^{N_d} \mathcal{P}(m) K_{\rm cl}(m/N_d)$, and reduces to the simple expression $K = 2 N_d^{-1} \left[ \avg{m} -N_d^{-1} \avg{m^2} \right] = 2p(1-p)(1-N_d^{-1})$, where $\avg{\cdot}$ indicates an average of a random variable following a binomial distribution. In particular, it can be seen that the maximum loss $K^{\rm max}=(1-N_d^{-1})/2$ monotonically decreases as a function of decreasing number of independent blockade regions, reaching $K=0$ uniformly for $N_d=1$. On the other hand, the maximum loss for any $N_d>1$ is achieved when $p=0.5$, which corresponds to a Rabi frequency of $\Omega^{\rm max} = \sqrt{\Gamma^2 + 4(\delta-\Delta_{\kpar=0})^2}/\sqrt{8 N_b}$.

\begin{figure}[t]
\begin{minipage}{0.4\textwidth}
\centering
\includegraphics[width=\textwidth]{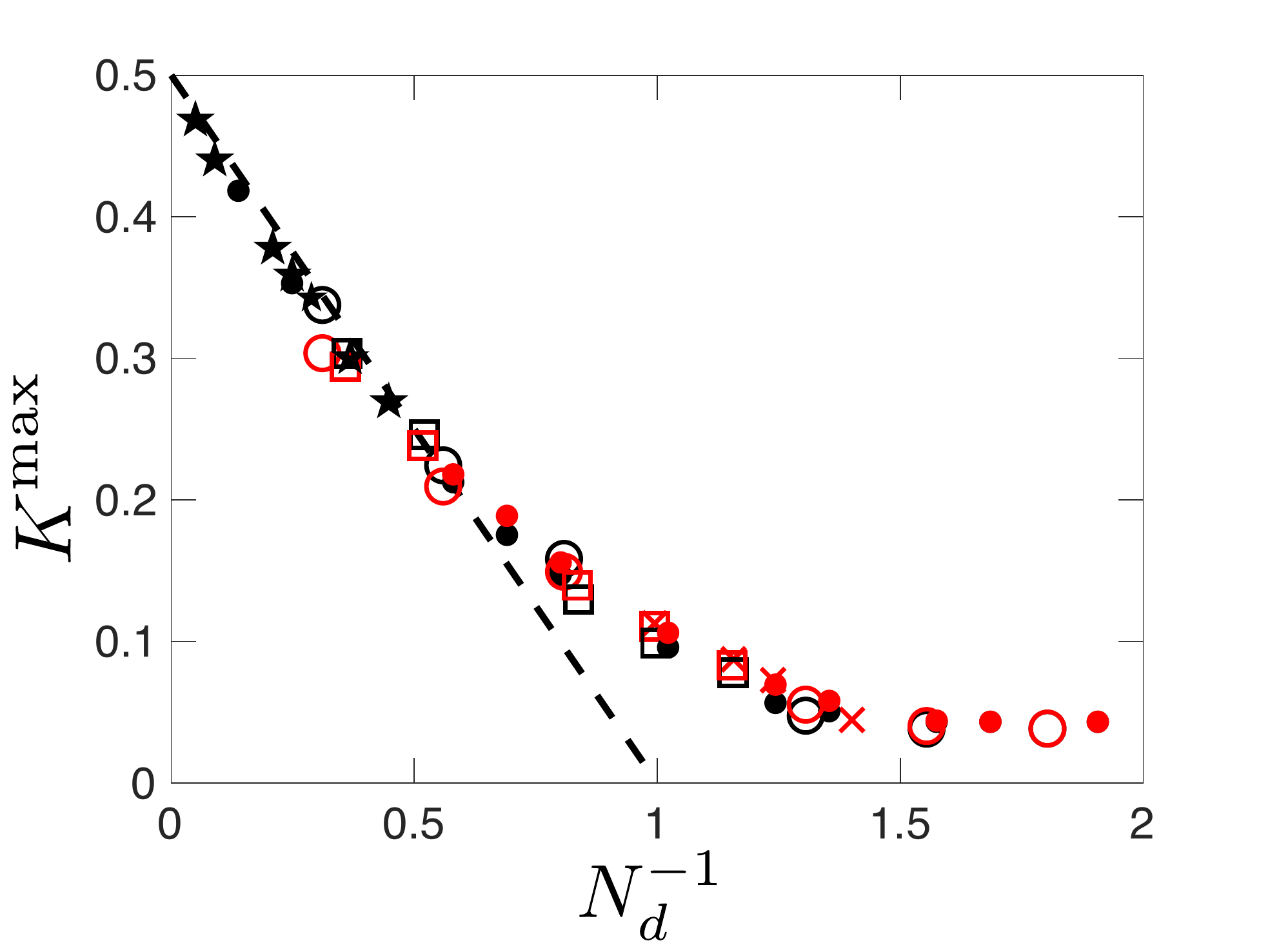}
\end{minipage}
\caption{Maximum loss as a function of $N_d^{-1} = N_b d^2/2\pi w_0^2$, where $N_d$ roughly corresponds to the number of blockade regions. Red and black symbols correspond to the exact numerical result and the numerical stochastic model, respectively, for an array with $N_a=16$ (open circles), $N_a=25$ (squares), $N_a=36$ (solid circles), $N_a=49$ (crosses) and $N_a=100$ (stars) atoms. The data collapse onto a universal curve that tends to the analytical model result $\Kmax = (1-N_d^{-1})/2$~(dashed line), in the limit of small $N_d^{-1}$. The simulations are done with a lattice constant of $d/\lambda_0 = 0.5$ and beam waist $w_0 = 0.4 \sqrt{N_a} d$).}
\label{FigS3c}
\end{figure}

In Fig.~\ref{FigS3c}, we compare the analytical approximation for $K^{\rm max}$ with numerical simulations, for square arrays with $N_a=16,25,36,49$, and $100$ atoms, and varying blockade radii $R_b$. These simulations involve both full numerical density matrix simulations~(red points), and the semi-classical stochastic model~(black). In particular, for each system size and blockade radius, we drive the system with a Gaussian beam of waist $w_0=0.4\sqrt{N_a}d$, and find the power at which the maximum loss $\Kmax$ occurs. We then plot $\Kmax$ as a function of the approximate number of blockade regions, $N_d= 2 \pi w_0^2/N_b d^2$. We find that the data points collapse onto a single universal curve, and that as long as $N_d^{-1}\lesssim 1$~(the regime of multiple independent blockade regions), this curve agrees well with the formula $\Kmax = (1-N_d^{-1})/2$~(dashed line). For large $N_d^{-1}$, the maximum loss becomes nearly zero, with the difference from the simple formula largely being attributable to the inhomogeneity of the beam across the system, which is not accounted for in our simple analytical model.

\newpage

\bibliography{Paper_RydbergArray.bib}

\begin{thebibliography}{56}%
\makeatletter
\providecommand \@ifxundefined [1]{%
 \@ifx{#1\undefined}
}%
\providecommand \@ifnum [1]{%
 \ifnum #1\expandafter \@firstoftwo
 \else \expandafter \@secondoftwo
 \fi
}%
\providecommand \@ifx [1]{%
 \ifx #1\expandafter \@firstoftwo
 \else \expandafter \@secondoftwo
 \fi
}%
\providecommand \natexlab [1]{#1}%
\providecommand \enquote  [1]{``#1''}%
\providecommand \bibnamefont  [1]{#1}%
\providecommand \bibfnamefont [1]{#1}%
\providecommand \citenamefont [1]{#1}%
\providecommand \href@noop [0]{\@secondoftwo}%
\providecommand \href [0]{\begingroup \@sanitize@url \@href}%
\providecommand \@href[1]{\@@startlink{#1}\@@href}%
\providecommand \@@href[1]{\endgroup#1\@@endlink}%
\providecommand \@sanitize@url [0]{\catcode `\\12\catcode `\$12\catcode
  `\&12\catcode `\#12\catcode `\^12\catcode `\_12\catcode `\%12\relax}%
\providecommand \@@startlink[1]{}%
\providecommand \@@endlink[0]{}%
\providecommand \url  [0]{\begingroup\@sanitize@url \@url }%
\providecommand \@url [1]{\endgroup\@href {#1}{\urlprefix }}%
\providecommand \urlprefix  [0]{URL }%
\providecommand \Eprint [0]{\href }%
\providecommand \doibase [0]{http://dx.doi.org/}%
\providecommand \selectlanguage [0]{\@gobble}%
\providecommand \bibinfo  [0]{\@secondoftwo}%
\providecommand \bibfield  [0]{\@secondoftwo}%
\providecommand \translation [1]{[#1]}%
\providecommand \BibitemOpen [0]{}%
\providecommand \bibitemStop [0]{}%
\providecommand \bibitemNoStop [0]{.\EOS\space}%
\providecommand \EOS [0]{\spacefactor3000\relax}%
\providecommand \BibitemShut  [1]{\csname bibitem#1\endcsname}%
\let\auto@bib@innerbib\@empty
\bibitem [{\citenamefont {Chang}\ \emph {et~al.}(2014)\citenamefont {Chang},
  \citenamefont {Vuleti{\'{c}}},\ and\ \citenamefont {Lukin}}]{ChVL14}%
  \BibitemOpen
  \bibfield  {author} {\bibinfo {author} {\bibfnamefont {D.~E.}\ \bibnamefont
  {Chang}}, \bibinfo {author} {\bibfnamefont {V.}~\bibnamefont
  {Vuleti{\'{c}}}}, \ and\ \bibinfo {author} {\bibfnamefont {M.~D.}\
  \bibnamefont {Lukin}},\ }\href {\doibase 10.1038/nphoton.2014.192} {\bibfield
   {journal} {\bibinfo  {journal} {Nature Photonics}\ }\textbf {\bibinfo
  {volume} {8}},\ \bibinfo {pages} {685} (\bibinfo {year} {2014})}\BibitemShut
  {NoStop}%
\bibitem [{\citenamefont {Fleischhauer}\ \emph {et~al.}(2005)\citenamefont
  {Fleischhauer}, \citenamefont {Imamoglu},\ and\ \citenamefont
  {Marangos}}]{FIM05}%
  \BibitemOpen
  \bibfield  {author} {\bibinfo {author} {\bibfnamefont {M.}~\bibnamefont
  {Fleischhauer}}, \bibinfo {author} {\bibfnamefont {A.}~\bibnamefont
  {Imamoglu}}, \ and\ \bibinfo {author} {\bibfnamefont {J.~P.}\ \bibnamefont
  {Marangos}},\ }\href {https://link.aps.org/doi/10.1103/RevModPhys.77.633}
  {\bibfield  {journal} {\bibinfo  {journal} {Rev. Mod. Phys.}\ }\textbf
  {\bibinfo {volume} {77}} (\bibinfo {year} {2005})}\BibitemShut {NoStop}%
\bibitem [{\citenamefont {Phillips}\ \emph {et~al.}(2001)\citenamefont
  {Phillips}, \citenamefont {Fleischhauer}, \citenamefont {Mair}, \citenamefont
  {Walsworth},\ and\ \citenamefont {Lukin}}]{PFM01}%
  \BibitemOpen
  \bibfield  {author} {\bibinfo {author} {\bibfnamefont {D.~F.}\ \bibnamefont
  {Phillips}}, \bibinfo {author} {\bibfnamefont {A.}~\bibnamefont
  {Fleischhauer}}, \bibinfo {author} {\bibfnamefont {A.}~\bibnamefont {Mair}},
  \bibinfo {author} {\bibfnamefont {R.~L.}\ \bibnamefont {Walsworth}}, \ and\
  \bibinfo {author} {\bibfnamefont {M.~D.}\ \bibnamefont {Lukin}},\ }\href
  {\doibase 10.1103/PhysRevLett.86.783} {\bibfield  {journal} {\bibinfo
  {journal} {Phys. Rev. Lett.}\ }\textbf {\bibinfo {volume} {86}},\ \bibinfo
  {pages} {783} (\bibinfo {year} {2001})}\BibitemShut {NoStop}%
\bibitem [{\citenamefont {Liu}\ \emph {et~al.}(2001)\citenamefont {Liu},
  \citenamefont {Dutton}, \citenamefont {Behroozi},\ and\ \citenamefont
  {Hau}}]{LDB01}%
  \BibitemOpen
  \bibfield  {author} {\bibinfo {author} {\bibfnamefont {C.}~\bibnamefont
  {Liu}}, \bibinfo {author} {\bibfnamefont {Z.}~\bibnamefont {Dutton}},
  \bibinfo {author} {\bibfnamefont {C.~H.}\ \bibnamefont {Behroozi}}, \ and\
  \bibinfo {author} {\bibfnamefont {L.~V.}\ \bibnamefont {Hau}},\ }\href
  {http://dx.doi.org/10.1038/35054017} {\bibfield  {journal} {\bibinfo
  {journal} {Nature}\ }\textbf {\bibinfo {volume} {409}},\ \bibinfo {pages}
  {490} (\bibinfo {year} {2001})}\BibitemShut {NoStop}%
\bibitem [{\citenamefont {Firstenberg}\ \emph {et~al.}(2013)\citenamefont
  {Firstenberg}, \citenamefont {Peyronel}, \citenamefont {Liang}, \citenamefont
  {Gorshkov}, \citenamefont {Lukin},\ and\ \citenamefont {Vuleti\'c}}]{FPL13}%
  \BibitemOpen
  \bibfield  {author} {\bibinfo {author} {\bibfnamefont {O.}~\bibnamefont
  {Firstenberg}}, \bibinfo {author} {\bibfnamefont {T.}~\bibnamefont
  {Peyronel}}, \bibinfo {author} {\bibfnamefont {Q.-Y.}\ \bibnamefont {Liang}},
  \bibinfo {author} {\bibfnamefont {A.~V.}\ \bibnamefont {Gorshkov}}, \bibinfo
  {author} {\bibfnamefont {M.~D.}\ \bibnamefont {Lukin}}, \ and\ \bibinfo
  {author} {\bibfnamefont {V.}~\bibnamefont {Vuleti\'c}},\ }\href
  {https://doi.org/10.1038/nature12512} {\bibfield  {journal} {\bibinfo
  {journal} {Nature}\ }\textbf {\bibinfo {volume} {502}},\ \bibinfo {pages}
  {71} (\bibinfo {year} {2013})}\BibitemShut {NoStop}%
\bibitem [{\citenamefont {Baur}\ \emph {et~al.}(2014)\citenamefont {Baur},
  \citenamefont {Tiarks}, \citenamefont {Rempe},\ and\ \citenamefont
  {D\"urr}}]{BTR14}%
  \BibitemOpen
  \bibfield  {author} {\bibinfo {author} {\bibfnamefont {S.}~\bibnamefont
  {Baur}}, \bibinfo {author} {\bibfnamefont {D.}~\bibnamefont {Tiarks}},
  \bibinfo {author} {\bibfnamefont {G.}~\bibnamefont {Rempe}}, \ and\ \bibinfo
  {author} {\bibfnamefont {S.}~\bibnamefont {D\"urr}},\ }\href {\doibase
  10.1103/PhysRevLett.112.073901} {\bibfield  {journal} {\bibinfo  {journal}
  {Phys. Rev. Lett.}\ }\textbf {\bibinfo {volume} {112}},\ \bibinfo {pages}
  {073901} (\bibinfo {year} {2014})}\BibitemShut {NoStop}%
\bibitem [{\citenamefont {Thompson}\ \emph {et~al.}(2017)\citenamefont
  {Thompson}, \citenamefont {Nicholson}, \citenamefont {Liang}, \citenamefont
  {Cantu}, \citenamefont {Venkatramani}, \citenamefont {Choi}, \citenamefont
  {Fedorov}, \citenamefont {Viscor}, \citenamefont {Pohl}, \citenamefont
  {Lukin},\ and\ \citenamefont {Vuletić}}]{TNL17}%
  \BibitemOpen
  \bibfield  {author} {\bibinfo {author} {\bibfnamefont {J.~D.}\ \bibnamefont
  {Thompson}}, \bibinfo {author} {\bibfnamefont {T.~L.}\ \bibnamefont
  {Nicholson}}, \bibinfo {author} {\bibfnamefont {Q.-Y.}\ \bibnamefont
  {Liang}}, \bibinfo {author} {\bibfnamefont {S.~H.}\ \bibnamefont {Cantu}},
  \bibinfo {author} {\bibfnamefont {A.~V.}\ \bibnamefont {Venkatramani}},
  \bibinfo {author} {\bibfnamefont {S.}~\bibnamefont {Choi}}, \bibinfo {author}
  {\bibfnamefont {I.~A.}\ \bibnamefont {Fedorov}}, \bibinfo {author}
  {\bibfnamefont {D.}~\bibnamefont {Viscor}}, \bibinfo {author} {\bibfnamefont
  {T.}~\bibnamefont {Pohl}}, \bibinfo {author} {\bibfnamefont {M.~D.}\
  \bibnamefont {Lukin}}, \ and\ \bibinfo {author} {\bibfnamefont
  {V.}~\bibnamefont {Vuletić}},\ }\href {https://doi.org/10.1038/nature20823}
  {\bibfield  {journal} {\bibinfo  {journal} {Nature}\ }\textbf {\bibinfo
  {volume} {542}},\ \bibinfo {pages} {206} (\bibinfo {year}
  {2017})}\BibitemShut {NoStop}%
\bibitem [{\citenamefont {Tiarks}\ \emph {et~al.}(2019)\citenamefont {Tiarks},
  \citenamefont {Schmidt-Eberle}, \citenamefont {Stolz}, \citenamefont
  {Rempe},\ and\ \citenamefont {D{\"u}rr}}]{ref:GateRempe19}%
  \BibitemOpen
  \bibfield  {author} {\bibinfo {author} {\bibfnamefont {D.}~\bibnamefont
  {Tiarks}}, \bibinfo {author} {\bibfnamefont {S.}~\bibnamefont
  {Schmidt-Eberle}}, \bibinfo {author} {\bibfnamefont {T.}~\bibnamefont
  {Stolz}}, \bibinfo {author} {\bibfnamefont {G.}~\bibnamefont {Rempe}}, \ and\
  \bibinfo {author} {\bibfnamefont {S.}~\bibnamefont {D{\"u}rr}},\ }\href
  {\doibase 10.1038/s41567-018-0313-7} {\bibfield  {journal} {\bibinfo
  {journal} {Nature Physics}\ }\textbf {\bibinfo {volume} {15}},\ \bibinfo
  {pages} {124} (\bibinfo {year} {2019})}\BibitemShut {NoStop}%
\bibitem [{\citenamefont {Pritchard}\ \emph {et~al.}(2010)\citenamefont
  {Pritchard}, \citenamefont {Maxwell}, \citenamefont {Gauguet}, \citenamefont
  {Weatherill}, \citenamefont {Jones},\ and\ \citenamefont
  {Adams}}]{ref:CSAdams10}%
  \BibitemOpen
  \bibfield  {author} {\bibinfo {author} {\bibfnamefont {J.~D.}\ \bibnamefont
  {Pritchard}}, \bibinfo {author} {\bibfnamefont {D.}~\bibnamefont {Maxwell}},
  \bibinfo {author} {\bibfnamefont {A.}~\bibnamefont {Gauguet}}, \bibinfo
  {author} {\bibfnamefont {K.~J.}\ \bibnamefont {Weatherill}}, \bibinfo
  {author} {\bibfnamefont {M.~P.~A.}\ \bibnamefont {Jones}}, \ and\ \bibinfo
  {author} {\bibfnamefont {C.~S.}\ \bibnamefont {Adams}},\ }\href {\doibase
  10.1103/PhysRevLett.105.193603} {\bibfield  {journal} {\bibinfo  {journal}
  {Phys. Rev. Lett.}\ }\textbf {\bibinfo {volume} {105}},\ \bibinfo {pages}
  {193603} (\bibinfo {year} {2010})}\BibitemShut {NoStop}%
\bibitem [{\citenamefont {Gorniaczyk}\ \emph {et~al.}(2014)\citenamefont
  {Gorniaczyk}, \citenamefont {Tresp}, \citenamefont {Schmidt}, \citenamefont
  {Fedder},\ and\ \citenamefont {Hofferberth}}]{ref:PRL113}%
  \BibitemOpen
  \bibfield  {author} {\bibinfo {author} {\bibfnamefont {H.}~\bibnamefont
  {Gorniaczyk}}, \bibinfo {author} {\bibfnamefont {C.}~\bibnamefont {Tresp}},
  \bibinfo {author} {\bibfnamefont {J.}~\bibnamefont {Schmidt}}, \bibinfo
  {author} {\bibfnamefont {H.}~\bibnamefont {Fedder}}, \ and\ \bibinfo {author}
  {\bibfnamefont {S.}~\bibnamefont {Hofferberth}},\ }\href {\doibase
  10.1103/PhysRevLett.113.053601} {\bibfield  {journal} {\bibinfo  {journal}
  {Phys. Rev. Lett.}\ }\textbf {\bibinfo {volume} {113}},\ \bibinfo {pages}
  {053601} (\bibinfo {year} {2014})}\BibitemShut {NoStop}%
\bibitem [{\citenamefont {Gorshkov}\ \emph {et~al.}(2011)\citenamefont
  {Gorshkov}, \citenamefont {Otterbach}, \citenamefont {Fleischhauer},
  \citenamefont {Pohl},\ and\ \citenamefont {Lukin}}]{ref:Gorshkov11}%
  \BibitemOpen
  \bibfield  {author} {\bibinfo {author} {\bibfnamefont {A.~V.}\ \bibnamefont
  {Gorshkov}}, \bibinfo {author} {\bibfnamefont {J.}~\bibnamefont {Otterbach}},
  \bibinfo {author} {\bibfnamefont {M.}~\bibnamefont {Fleischhauer}}, \bibinfo
  {author} {\bibfnamefont {T.}~\bibnamefont {Pohl}}, \ and\ \bibinfo {author}
  {\bibfnamefont {M.~D.}\ \bibnamefont {Lukin}},\ }\href {\doibase
  10.1103/PhysRevLett.107.133602} {\bibfield  {journal} {\bibinfo  {journal}
  {Phys. Rev. Lett.}\ }\textbf {\bibinfo {volume} {107}},\ \bibinfo {pages}
  {133602} (\bibinfo {year} {2011})}\BibitemShut {NoStop}%
\bibitem [{\citenamefont {Zoubi}\ and\ \citenamefont {Ritsch}(2011)}]{ZR11}%
  \BibitemOpen
  \bibfield  {author} {\bibinfo {author} {\bibfnamefont {H.}~\bibnamefont
  {Zoubi}}\ and\ \bibinfo {author} {\bibfnamefont {H.}~\bibnamefont {Ritsch}},\
  }\href@noop {} {\bibfield  {journal} {\bibinfo  {journal} {Phys. Rev. A}\
  }\textbf {\bibinfo {volume} {83}},\ \bibinfo {pages} {063831} (\bibinfo
  {year} {2011})}\BibitemShut {NoStop}%
\bibitem [{\citenamefont {Jenkins}\ and\ \citenamefont
  {Ruostekoski}(2012)}]{JR12}%
  \BibitemOpen
  \bibfield  {author} {\bibinfo {author} {\bibfnamefont {S.~D.}\ \bibnamefont
  {Jenkins}}\ and\ \bibinfo {author} {\bibfnamefont {J.}~\bibnamefont
  {Ruostekoski}},\ }\href@noop {} {\bibfield  {journal} {\bibinfo  {journal}
  {Phys. Rev. A}\ }\textbf {\bibinfo {volume} {86}},\ \bibinfo {pages} {031602}
  (\bibinfo {year} {2012})}\BibitemShut {NoStop}%
\bibitem [{\citenamefont {Plankensteiner}\ \emph {et~al.}(2015)\citenamefont
  {Plankensteiner}, \citenamefont {Ostermann}, \citenamefont {Ritsch},\ and\
  \citenamefont {Genes}}]{POR15}%
  \BibitemOpen
  \bibfield  {author} {\bibinfo {author} {\bibfnamefont {D.}~\bibnamefont
  {Plankensteiner}}, \bibinfo {author} {\bibfnamefont {L.}~\bibnamefont
  {Ostermann}}, \bibinfo {author} {\bibfnamefont {H.}~\bibnamefont {Ritsch}}, \
  and\ \bibinfo {author} {\bibfnamefont {C.}~\bibnamefont {Genes}},\
  }\href@noop {} {\bibfield  {journal} {\bibinfo  {journal} {Sci.\ Rep.}\
  }\textbf {\bibinfo {volume} {5}},\ \bibinfo {pages} {16231} (\bibinfo {year}
  {2015})}\BibitemShut {NoStop}%
\bibitem [{\citenamefont {Bettles}\ \emph {et~al.}(2015)\citenamefont
  {Bettles}, \citenamefont {Gardiner},\ and\ \citenamefont {Adams}}]{BGA15}%
  \BibitemOpen
  \bibfield  {author} {\bibinfo {author} {\bibfnamefont {R.~J.}\ \bibnamefont
  {Bettles}}, \bibinfo {author} {\bibfnamefont {S.~A.}\ \bibnamefont
  {Gardiner}}, \ and\ \bibinfo {author} {\bibfnamefont {C.~S.}\ \bibnamefont
  {Adams}},\ }\href@noop {} {\bibfield  {journal} {\bibinfo  {journal} {Phys.
  Rev. A}\ }\textbf {\bibinfo {volume} {92}},\ \bibinfo {pages} {063822}
  (\bibinfo {year} {2015})}\BibitemShut {NoStop}%
\bibitem [{\citenamefont {Bettles}\ \emph
  {et~al.}(2016{\natexlab{a}})\citenamefont {Bettles}, \citenamefont
  {Gardiner},\ and\ \citenamefont {Adams}}]{BGA16}%
  \BibitemOpen
  \bibfield  {author} {\bibinfo {author} {\bibfnamefont {R.~J.}\ \bibnamefont
  {Bettles}}, \bibinfo {author} {\bibfnamefont {S.~A.}\ \bibnamefont
  {Gardiner}}, \ and\ \bibinfo {author} {\bibfnamefont {C.~S.}\ \bibnamefont
  {Adams}},\ }\href@noop {} {\bibfield  {journal} {\bibinfo  {journal} {Phys.
  Rev. A}\ }\textbf {\bibinfo {volume} {94}},\ \bibinfo {pages} {043844}
  (\bibinfo {year} {2016}{\natexlab{a}})}\BibitemShut {NoStop}%
\bibitem [{\citenamefont {Bettles}\ \emph
  {et~al.}(2016{\natexlab{b}})\citenamefont {Bettles}, \citenamefont
  {Gardiner},\ and\ \citenamefont {Adams}}]{BGA16b}%
  \BibitemOpen
  \bibfield  {author} {\bibinfo {author} {\bibfnamefont {R.~J.}\ \bibnamefont
  {Bettles}}, \bibinfo {author} {\bibfnamefont {S.~A.}\ \bibnamefont
  {Gardiner}}, \ and\ \bibinfo {author} {\bibfnamefont {C.~S.}\ \bibnamefont
  {Adams}},\ }\href@noop {} {\bibfield  {journal} {\bibinfo  {journal} {Phys.
  Rev. Lett.}\ }\textbf {\bibinfo {volume} {116}},\ \bibinfo {pages} {103602}
  (\bibinfo {year} {2016}{\natexlab{b}})}\BibitemShut {NoStop}%
\bibitem [{\citenamefont {Sutherland}\ and\ \citenamefont
  {Robicheaux}(2016)}]{SR16}%
  \BibitemOpen
  \bibfield  {author} {\bibinfo {author} {\bibfnamefont {R.~T.}\ \bibnamefont
  {Sutherland}}\ and\ \bibinfo {author} {\bibfnamefont {F.}~\bibnamefont
  {Robicheaux}},\ }\href@noop {} {\bibfield  {journal} {\bibinfo  {journal}
  {PRA}\ }\textbf {\bibinfo {volume} {94}},\ \bibinfo {pages} {013847}
  (\bibinfo {year} {2016})}\BibitemShut {NoStop}%
\bibitem [{\citenamefont {Shahmoon}\ \emph {et~al.}(2017)\citenamefont
  {Shahmoon}, \citenamefont {Wild}, \citenamefont {Lukin},\ and\ \citenamefont
  {Yelin}}]{SWL17}%
  \BibitemOpen
  \bibfield  {author} {\bibinfo {author} {\bibfnamefont {E.}~\bibnamefont
  {Shahmoon}}, \bibinfo {author} {\bibfnamefont {D.~S.}\ \bibnamefont {Wild}},
  \bibinfo {author} {\bibfnamefont {M.~D.}\ \bibnamefont {Lukin}}, \ and\
  \bibinfo {author} {\bibfnamefont {S.~F.}\ \bibnamefont {Yelin}},\ }\href@noop
  {} {\bibfield  {journal} {\bibinfo  {journal} {Phys. Rev. Lett.}\ }\textbf
  {\bibinfo {volume} {118}},\ \bibinfo {pages} {113601} (\bibinfo {year}
  {2017})}\BibitemShut {NoStop}%
\bibitem [{\citenamefont {Perczel}\ \emph {et~al.}(2017)\citenamefont
  {Perczel}, \citenamefont {Borregaard}, \citenamefont {Chang}, \citenamefont
  {Pichler}, \citenamefont {Yelin}, \citenamefont {Zoller},\ and\ \citenamefont
  {Lukin}}]{PBCh17}%
  \BibitemOpen
  \bibfield  {author} {\bibinfo {author} {\bibfnamefont {J.}~\bibnamefont
  {Perczel}}, \bibinfo {author} {\bibfnamefont {J.}~\bibnamefont {Borregaard}},
  \bibinfo {author} {\bibfnamefont {D.~E.}\ \bibnamefont {Chang}}, \bibinfo
  {author} {\bibfnamefont {H.}~\bibnamefont {Pichler}}, \bibinfo {author}
  {\bibfnamefont {S.~F.}\ \bibnamefont {Yelin}}, \bibinfo {author}
  {\bibfnamefont {P.}~\bibnamefont {Zoller}}, \ and\ \bibinfo {author}
  {\bibfnamefont {M.~D.}\ \bibnamefont {Lukin}},\ }\href {\doibase
  10.1103/PhysRevLett.119.023603} {\bibfield  {journal} {\bibinfo  {journal}
  {Phys. Rev. Lett.}\ }\textbf {\bibinfo {volume} {119}},\ \bibinfo {pages}
  {023603} (\bibinfo {year} {2017})}\BibitemShut {NoStop}%
\bibitem [{\citenamefont {Manzoni}\ \emph {et~al.}(2018)\citenamefont
  {Manzoni}, \citenamefont {Moreno-Cardoner}, \citenamefont {Asenjo-Garcia},
  \citenamefont {Porto}, \citenamefont {Gorshkov},\ and\ \citenamefont
  {Chang}}]{MMA18}%
  \BibitemOpen
  \bibfield  {author} {\bibinfo {author} {\bibfnamefont {M.~T.}\ \bibnamefont
  {Manzoni}}, \bibinfo {author} {\bibfnamefont {M.}~\bibnamefont
  {Moreno-Cardoner}}, \bibinfo {author} {\bibfnamefont {A.}~\bibnamefont
  {Asenjo-Garcia}}, \bibinfo {author} {\bibfnamefont {J.~V.}\ \bibnamefont
  {Porto}}, \bibinfo {author} {\bibfnamefont {A.~V.}\ \bibnamefont {Gorshkov}},
  \ and\ \bibinfo {author} {\bibfnamefont {D.~E.}\ \bibnamefont {Chang}},\
  }\href@noop {} {\bibfield  {journal} {\bibinfo  {journal} {New Journal of
  Physics}\ }\textbf {\bibinfo {volume} {20}},\ \bibinfo {pages} {083048}
  (\bibinfo {year} {2018})}\BibitemShut {NoStop}%
\bibitem [{\citenamefont {{Pi{\~{n}}eiro Orioli}}\ and\ \citenamefont
  {Rey}(2019)}]{PR19}%
  \BibitemOpen
  \bibfield  {author} {\bibinfo {author} {\bibfnamefont {A.}~\bibnamefont
  {{Pi{\~{n}}eiro Orioli}}}\ and\ \bibinfo {author} {\bibfnamefont {A.~M.}\
  \bibnamefont {Rey}},\ }\href {https://doi.org/10.1103/PhysRevLett.123.223601}
  {\bibfield  {journal} {\bibinfo  {journal} {Physical Review Letters}\
  }\textbf {\bibinfo {volume} {123}},\ \bibinfo {pages} {223601} (\bibinfo
  {year} {2019})}\BibitemShut {NoStop}%
\bibitem [{\citenamefont {Rui}\ \emph {et~al.}(2020)\citenamefont {Rui},
  \citenamefont {Wei}, \citenamefont {Rubio-Abadal}, \citenamefont {Hollerith},
  \citenamefont {Zeiher}, \citenamefont {Stamper-Kurn}, \citenamefont {Gross},\
  and\ \citenamefont {Bloch}}]{RWR20}%
  \BibitemOpen
  \bibfield  {author} {\bibinfo {author} {\bibfnamefont {J.}~\bibnamefont
  {Rui}}, \bibinfo {author} {\bibfnamefont {D.}~\bibnamefont {Wei}}, \bibinfo
  {author} {\bibfnamefont {A.}~\bibnamefont {Rubio-Abadal}}, \bibinfo {author}
  {\bibfnamefont {S.}~\bibnamefont {Hollerith}}, \bibinfo {author}
  {\bibfnamefont {J.}~\bibnamefont {Zeiher}}, \bibinfo {author} {\bibfnamefont
  {D.~M.}\ \bibnamefont {Stamper-Kurn}}, \bibinfo {author} {\bibfnamefont
  {C.}~\bibnamefont {Gross}}, \ and\ \bibinfo {author} {\bibfnamefont
  {I.}~\bibnamefont {Bloch}},\ }\href
  {https://doi.org/10.1038/s41586-020-2463-x} {\bibfield  {journal} {\bibinfo
  {journal} {Nature}\ }\textbf {\bibinfo {volume} {583}},\ \bibinfo {pages}
  {369} (\bibinfo {year} {2020})}\BibitemShut {NoStop}%
\bibitem [{\citenamefont {Zhang}\ and\ \citenamefont {M{\o}lmer}(2019)}]{ZM19}%
  \BibitemOpen
  \bibfield  {author} {\bibinfo {author} {\bibfnamefont {Y.~X.}\ \bibnamefont
  {Zhang}}\ and\ \bibinfo {author} {\bibfnamefont {K.}~\bibnamefont
  {M{\o}lmer}},\ }\href {https://doi.org/10.1103/PhysRevLett.122.203605}
  {\bibfield  {journal} {\bibinfo  {journal} {Physical Review Letters}\
  }\textbf {\bibinfo {volume} {122}},\ \bibinfo {pages} {203605} (\bibinfo
  {year} {2019})}\BibitemShut {NoStop}%
\bibitem [{\citenamefont {Zhang}\ \emph {et~al.}(2020)\citenamefont {Zhang},
  \citenamefont {Yu},\ and\ \citenamefont {M{\o}lmer}}]{ZChM20}%
  \BibitemOpen
  \bibfield  {author} {\bibinfo {author} {\bibfnamefont {Y.-X.}\ \bibnamefont
  {Zhang}}, \bibinfo {author} {\bibfnamefont {C.}~\bibnamefont {Yu}}, \ and\
  \bibinfo {author} {\bibfnamefont {K.}~\bibnamefont {M{\o}lmer}},\ }\href@noop
  {} {\bibfield  {journal} {\bibinfo  {journal} {Physical Review Research}\
  }\textbf {\bibinfo {volume} {2}},\ \bibinfo {pages} {1} (\bibinfo {year}
  {2020})}\BibitemShut {NoStop}%
\bibitem [{\citenamefont {{Pi{\~{n}}eiro Orioli}}\ and\ \citenamefont
  {Rey}(2020)}]{PR20}%
  \BibitemOpen
  \bibfield  {author} {\bibinfo {author} {\bibfnamefont {A.}~\bibnamefont
  {{Pi{\~{n}}eiro Orioli}}}\ and\ \bibinfo {author} {\bibfnamefont {A.~M.}\
  \bibnamefont {Rey}},\ }\href@noop {} {\bibfield  {journal} {\bibinfo
  {journal} {Physical Review A}\ }\textbf {\bibinfo {volume} {101}},\ \bibinfo
  {pages} {043816} (\bibinfo {year} {2020})}\BibitemShut {NoStop}%
\bibitem [{\citenamefont {Bekenstein}\ \emph {et~al.}(2020)\citenamefont
  {Bekenstein}, \citenamefont {Pikovski}, \citenamefont {Pichler},
  \citenamefont {Shahmoon}, \citenamefont {Yelin},\ and\ \citenamefont
  {Lukin}}]{ref:QuantumMetasurfaces}%
  \BibitemOpen
  \bibfield  {author} {\bibinfo {author} {\bibfnamefont {R.}~\bibnamefont
  {Bekenstein}}, \bibinfo {author} {\bibfnamefont {I.}~\bibnamefont
  {Pikovski}}, \bibinfo {author} {\bibfnamefont {H.}~\bibnamefont {Pichler}},
  \bibinfo {author} {\bibfnamefont {E.}~\bibnamefont {Shahmoon}}, \bibinfo
  {author} {\bibfnamefont {S.~F.}\ \bibnamefont {Yelin}}, \ and\ \bibinfo
  {author} {\bibfnamefont {M.~D.}\ \bibnamefont {Lukin}},\ }\href {\doibase
  10.1038/s41567-020-0845-5} {\bibfield  {journal} {\bibinfo  {journal} {Nature
  Physics}\ }\textbf {\bibinfo {volume} {16}},\ \bibinfo {pages} {676}
  (\bibinfo {year} {2020})}\BibitemShut {NoStop}%
\bibitem [{\citenamefont {Wei}\ \emph {et~al.}(2020)\citenamefont {Wei},
  \citenamefont {Malz}, \citenamefont {González-Tudela},\ and\ \citenamefont
  {Cirac}}]{ref:Cirac}%
  \BibitemOpen
  \bibfield  {author} {\bibinfo {author} {\bibfnamefont {Z.-Y.}\ \bibnamefont
  {Wei}}, \bibinfo {author} {\bibfnamefont {D.}~\bibnamefont {Malz}}, \bibinfo
  {author} {\bibfnamefont {A.}~\bibnamefont {González-Tudela}}, \ and\
  \bibinfo {author} {\bibfnamefont {J.~I.}\ \bibnamefont {Cirac}},\ }\href@noop
  {} {\bibfield  {journal} {\bibinfo  {journal} {ArXiv:2011.03919}\ } (\bibinfo
  {year} {2020})}\BibitemShut {NoStop}%
\bibitem [{\citenamefont {{De Abajo}}(2007)}]{DeA07}%
  \BibitemOpen
  \bibfield  {author} {\bibinfo {author} {\bibfnamefont {F.~J.}\ \bibnamefont
  {{De Abajo}}},\ }\href {\doibase 10.1103/RevModPhys.79.1267} {\bibfield
  {journal} {\bibinfo  {journal} {Reviews of Modern Physics}\ }\textbf
  {\bibinfo {volume} {79}},\ \bibinfo {pages} {1267} (\bibinfo {year}
  {2007})}\BibitemShut {NoStop}%
\bibitem [{\citenamefont {Bettles}\ \emph {et~al.}(2020)\citenamefont
  {Bettles}, \citenamefont {Lee}, \citenamefont {Gardiner},\ and\ \citenamefont
  {Ruostekoski}}]{ref:Bettles2020}%
  \BibitemOpen
  \bibfield  {author} {\bibinfo {author} {\bibfnamefont {R.~J.}\ \bibnamefont
  {Bettles}}, \bibinfo {author} {\bibfnamefont {M.~D.}\ \bibnamefont {Lee}},
  \bibinfo {author} {\bibfnamefont {S.~A.}\ \bibnamefont {Gardiner}}, \ and\
  \bibinfo {author} {\bibfnamefont {J.}~\bibnamefont {Ruostekoski}},\ }\href
  {\doibase 10.1038/s42005-020-00404-3} {\bibfield  {journal} {\bibinfo
  {journal} {Communications Physics}\ }\textbf {\bibinfo {volume} {3}},\
  \bibinfo {pages} {141} (\bibinfo {year} {2020})}\BibitemShut {NoStop}%
\bibitem [{\citenamefont {Cidrim}\ \emph {et~al.}(2020)\citenamefont {Cidrim},
  \citenamefont {do~Espirito~Santo}, \citenamefont {Schachenmayer},
  \citenamefont {Kaiser},\ and\ \citenamefont
  {Bachelard}}]{ref:Subwavelength1}%
  \BibitemOpen
  \bibfield  {author} {\bibinfo {author} {\bibfnamefont {A.}~\bibnamefont
  {Cidrim}}, \bibinfo {author} {\bibfnamefont {T.~S.}\ \bibnamefont
  {do~Espirito~Santo}}, \bibinfo {author} {\bibfnamefont {J.}~\bibnamefont
  {Schachenmayer}}, \bibinfo {author} {\bibfnamefont {R.}~\bibnamefont
  {Kaiser}}, \ and\ \bibinfo {author} {\bibfnamefont {R.}~\bibnamefont
  {Bachelard}},\ }\href {\doibase 10.1103/PhysRevLett.125.073601} {\bibfield
  {journal} {\bibinfo  {journal} {Phys. Rev. Lett.}\ }\textbf {\bibinfo
  {volume} {125}},\ \bibinfo {pages} {073601} (\bibinfo {year}
  {2020})}\BibitemShut {NoStop}%
\bibitem [{\citenamefont {Williamson}\ \emph {et~al.}(2020)\citenamefont
  {Williamson}, \citenamefont {Borgh},\ and\ \citenamefont
  {Ruostekoski}}]{ref:Subwavelength2}%
  \BibitemOpen
  \bibfield  {author} {\bibinfo {author} {\bibfnamefont {L.~A.}\ \bibnamefont
  {Williamson}}, \bibinfo {author} {\bibfnamefont {M.~O.}\ \bibnamefont
  {Borgh}}, \ and\ \bibinfo {author} {\bibfnamefont {J.}~\bibnamefont
  {Ruostekoski}},\ }\href {\doibase 10.1103/PhysRevLett.125.073602} {\bibfield
  {journal} {\bibinfo  {journal} {Phys. Rev. Lett.}\ }\textbf {\bibinfo
  {volume} {125}},\ \bibinfo {pages} {073602} (\bibinfo {year}
  {2020})}\BibitemShut {NoStop}%
\bibitem [{\citenamefont {Shahmoon}\ \emph {et~al.}(2020)\citenamefont
  {Shahmoon}, \citenamefont {Lukin},\ and\ \citenamefont {Yelin}}]{SLY20}%
  \BibitemOpen
  \bibfield  {author} {\bibinfo {author} {\bibfnamefont {E.}~\bibnamefont
  {Shahmoon}}, \bibinfo {author} {\bibfnamefont {M.~D.}\ \bibnamefont {Lukin}},
  \ and\ \bibinfo {author} {\bibfnamefont {S.~F.}\ \bibnamefont {Yelin}},\
  }\href {\doibase 10.1103/PhysRevA.101.063833} {\bibfield  {journal} {\bibinfo
   {journal} {Phys. Rev. A}\ }\textbf {\bibinfo {volume} {101}},\ \bibinfo
  {pages} {063833} (\bibinfo {year} {2020})}\BibitemShut {NoStop}%
\bibitem [{\citenamefont {Meystre}\ and\ \citenamefont
  {Sargent}(2007)}]{MS1990}%
  \BibitemOpen
  \bibfield  {author} {\bibinfo {author} {\bibfnamefont {P.}~\bibnamefont
  {Meystre}}\ and\ \bibinfo {author} {\bibfnamefont {M.}~\bibnamefont
  {Sargent}},\ }\href@noop {} {\emph {\bibinfo {title} {Elements of Quantum
  Optics}}}\ (\bibinfo  {publisher} {Springer-Verlag},\ \bibinfo {address}
  {Berlin},\ \bibinfo {year} {2007})\BibitemShut {NoStop}%
\bibitem [{\citenamefont {Gruner}\ and\ \citenamefont
  {Welsch}(1996)}]{Gruner1996}%
  \BibitemOpen
  \bibfield  {author} {\bibinfo {author} {\bibfnamefont {T.}~\bibnamefont
  {Gruner}}\ and\ \bibinfo {author} {\bibfnamefont {D.-G.}\ \bibnamefont
  {Welsch}},\ }\href {\doibase 10.1103/PhysRevA.53.1818} {\bibfield  {journal}
  {\bibinfo  {journal} {Phys. Rev. A}\ }\textbf {\bibinfo {volume} {53}},\
  \bibinfo {pages} {1818} (\bibinfo {year} {1996})}\BibitemShut {NoStop}%
\bibitem [{\citenamefont {Dung}\ \emph {et~al.}(2002)\citenamefont {Dung},
  \citenamefont {Kn\"oll},\ and\ \citenamefont {Welsch}}]{Dung2002}%
  \BibitemOpen
  \bibfield  {author} {\bibinfo {author} {\bibfnamefont {H.~T.}\ \bibnamefont
  {Dung}}, \bibinfo {author} {\bibfnamefont {L.}~\bibnamefont {Kn\"oll}}, \
  and\ \bibinfo {author} {\bibfnamefont {D.-G.}\ \bibnamefont {Welsch}},\
  }\href {\doibase 10.1103/PhysRevA.66.063810} {\bibfield  {journal} {\bibinfo
  {journal} {Phys. Rev. A}\ }\textbf {\bibinfo {volume} {66}},\ \bibinfo
  {pages} {063810} (\bibinfo {year} {2002})}\BibitemShut {NoStop}%
\bibitem [{\citenamefont {Buhmann}\ and\ \citenamefont
  {Welsch}(2007)}]{Buhmann2007}%
  \BibitemOpen
  \bibfield  {author} {\bibinfo {author} {\bibfnamefont {S.~Y.}\ \bibnamefont
  {Buhmann}}\ and\ \bibinfo {author} {\bibfnamefont {D.-G.}\ \bibnamefont
  {Welsch}},\ }\href@noop {} {\bibfield  {journal} {\bibinfo  {journal} {Progr.
  in Quant. Electron.}\ }\textbf {\bibinfo {volume} {31}},\ \bibinfo {pages}
  {51 } (\bibinfo {year} {2007})}\BibitemShut {NoStop}%
\bibitem [{\citenamefont {Asenjo-Garcia}\ \emph
  {et~al.}(2017{\natexlab{a}})\citenamefont {Asenjo-Garcia}, \citenamefont
  {Hood}, \citenamefont {Chang},\ and\ \citenamefont {Kimble}}]{AHCh17}%
  \BibitemOpen
  \bibfield  {author} {\bibinfo {author} {\bibfnamefont {A.}~\bibnamefont
  {Asenjo-Garcia}}, \bibinfo {author} {\bibfnamefont {J.~D.}\ \bibnamefont
  {Hood}}, \bibinfo {author} {\bibfnamefont {D.~E.}\ \bibnamefont {Chang}}, \
  and\ \bibinfo {author} {\bibfnamefont {H.~J.}\ \bibnamefont {Kimble}},\
  }\href {\doibase 10.1103/PhysRevA.95.033818} {\bibfield  {journal} {\bibinfo
  {journal} {Phys. Rev. A}\ }\textbf {\bibinfo {volume} {95}},\ \bibinfo
  {pages} {033818} (\bibinfo {year} {2017}{\natexlab{a}})}\BibitemShut
  {NoStop}%
\bibitem [{\citenamefont {Asenjo-Garcia}\ \emph
  {et~al.}(2017{\natexlab{b}})\citenamefont {Asenjo-Garcia}, \citenamefont
  {Moreno-Cardoner}, \citenamefont {Albrecht}, \citenamefont {Kimble},\ and\
  \citenamefont {Chang}}]{AMA2017}%
  \BibitemOpen
  \bibfield  {author} {\bibinfo {author} {\bibfnamefont {A.}~\bibnamefont
  {Asenjo-Garcia}}, \bibinfo {author} {\bibfnamefont {M.}~\bibnamefont
  {Moreno-Cardoner}}, \bibinfo {author} {\bibfnamefont {A.}~\bibnamefont
  {Albrecht}}, \bibinfo {author} {\bibfnamefont {H.~J.}\ \bibnamefont
  {Kimble}}, \ and\ \bibinfo {author} {\bibfnamefont {D.~E.}\ \bibnamefont
  {Chang}},\ }\href {\doibase 10.1103/PhysRevX.7.031024} {\bibfield  {journal}
  {\bibinfo  {journal} {Phys. Rev. X}\ }\textbf {\bibinfo {volume} {7}},\
  \bibinfo {pages} {031024} (\bibinfo {year} {2017}{\natexlab{b}})}\BibitemShut
  {NoStop}%
\bibitem [{\citenamefont {Zeiher}\ \emph {et~al.}(2016)\citenamefont {Zeiher},
  \citenamefont {van Bijnen}, \citenamefont {Schauß}, \citenamefont {Hild},
  \citenamefont {Choi}, \citenamefont {Pohl}, \citenamefont {Bloch},\ and\
  \citenamefont {Gross}}]{ZvBS16}%
  \BibitemOpen
  \bibfield  {author} {\bibinfo {author} {\bibfnamefont {J.}~\bibnamefont
  {Zeiher}}, \bibinfo {author} {\bibfnamefont {R.}~\bibnamefont {van Bijnen}},
  \bibinfo {author} {\bibfnamefont {P.}~\bibnamefont {Schauß}}, \bibinfo
  {author} {\bibfnamefont {S.}~\bibnamefont {Hild}}, \bibinfo {author}
  {\bibfnamefont {J.-y.}\ \bibnamefont {Choi}}, \bibinfo {author}
  {\bibfnamefont {T.}~\bibnamefont {Pohl}}, \bibinfo {author} {\bibfnamefont
  {I.}~\bibnamefont {Bloch}}, \ and\ \bibinfo {author} {\bibfnamefont
  {C.}~\bibnamefont {Gross}},\ }\href@noop {} {\bibfield  {journal} {\bibinfo
  {journal} {Nature Physics}\ }\textbf {\bibinfo {volume} {12}},\ \bibinfo
  {pages} {1095} (\bibinfo {year} {2016})}\BibitemShut {NoStop}%
\bibitem [{\citenamefont {Zeiher}\ \emph {et~al.}(2017)\citenamefont {Zeiher},
  \citenamefont {Choi}, \citenamefont {Rubio-Abadal}, \citenamefont {Pohl},
  \citenamefont {van Bijnen}, \citenamefont {Bloch},\ and\ \citenamefont
  {Gross}}]{ZChR17}%
  \BibitemOpen
  \bibfield  {author} {\bibinfo {author} {\bibfnamefont {J.}~\bibnamefont
  {Zeiher}}, \bibinfo {author} {\bibfnamefont {J.-y.}\ \bibnamefont {Choi}},
  \bibinfo {author} {\bibfnamefont {A.}~\bibnamefont {Rubio-Abadal}}, \bibinfo
  {author} {\bibfnamefont {T.}~\bibnamefont {Pohl}}, \bibinfo {author}
  {\bibfnamefont {R.}~\bibnamefont {van Bijnen}}, \bibinfo {author}
  {\bibfnamefont {I.}~\bibnamefont {Bloch}}, \ and\ \bibinfo {author}
  {\bibfnamefont {C.}~\bibnamefont {Gross}},\ }\href {\doibase
  10.1103/PhysRevX.7.041063} {\bibfield  {journal} {\bibinfo  {journal} {Phys.
  Rev. X}\ }\textbf {\bibinfo {volume} {7}},\ \bibinfo {pages} {041063}
  (\bibinfo {year} {2017})}\BibitemShut {NoStop}%
\bibitem [{\citenamefont {Henkel}\ \emph {et~al.}(2010)\citenamefont {Henkel},
  \citenamefont {Nath},\ and\ \citenamefont {Pohl}}]{Macri-pohl}%
  \BibitemOpen
  \bibfield  {author} {\bibinfo {author} {\bibfnamefont {N.}~\bibnamefont
  {Henkel}}, \bibinfo {author} {\bibfnamefont {R.}~\bibnamefont {Nath}}, \ and\
  \bibinfo {author} {\bibfnamefont {T.}~\bibnamefont {Pohl}},\ }\href {\doibase
  10.1103/PhysRevLett.104.195302} {\bibfield  {journal} {\bibinfo  {journal}
  {Phys. Rev. Lett.}\ }\textbf {\bibinfo {volume} {104}},\ \bibinfo {pages}
  {195302} (\bibinfo {year} {2010})}\BibitemShut {NoStop}%
\bibitem [{\citenamefont {Macr\`{\i}}\ and\ \citenamefont
  {Pohl}(2014)}]{Macri-pohl2}%
  \BibitemOpen
  \bibfield  {author} {\bibinfo {author} {\bibfnamefont {T.}~\bibnamefont
  {Macr\`{\i}}}\ and\ \bibinfo {author} {\bibfnamefont {T.}~\bibnamefont
  {Pohl}},\ }\href {\doibase 10.1103/PhysRevA.89.011402} {\bibfield  {journal}
  {\bibinfo  {journal} {Phys. Rev. A}\ }\textbf {\bibinfo {volume} {89}},\
  \bibinfo {pages} {011402} (\bibinfo {year} {2014})}\BibitemShut {NoStop}%
\bibitem [{\citenamefont {Ko{\'{s}}cik}\ and\ \citenamefont
  {Sowi{\'{n}}ski}(2019)}]{Koscik2019}%
  \BibitemOpen
  \bibfield  {author} {\bibinfo {author} {\bibfnamefont {P.}~\bibnamefont
  {Ko{\'{s}}cik}}\ and\ \bibinfo {author} {\bibfnamefont {T.}~\bibnamefont
  {Sowi{\'{n}}ski}},\ }\href {\doibase 10.1038/s41598-019-48442-4} {\bibfield
  {journal} {\bibinfo  {journal} {Scientific Reports}\ }\textbf {\bibinfo
  {volume} {9}},\ \bibinfo {pages} {12018} (\bibinfo {year}
  {2019})}\BibitemShut {NoStop}%
\bibitem [{\citenamefont {Meystre}\ and\ \citenamefont
  {Sargent}(1998)}]{ref:textbook}%
  \BibitemOpen
  \bibfield  {author} {\bibinfo {author} {\bibfnamefont {P.}~\bibnamefont
  {Meystre}}\ and\ \bibinfo {author} {\bibfnamefont {M.}~\bibnamefont
  {Sargent}},\ }\href {https://books.google.es/books?id=dWnIOHloxoEC} {\emph
  {\bibinfo {title} {Elements of Quantum Optics}}}\ (\bibinfo  {publisher}
  {Springer Berlin Heidelberg},\ \bibinfo {year} {1998})\BibitemShut {NoStop}%
\bibitem [{\citenamefont {Tanaka}\ \emph {et~al.}(1985)\citenamefont {Tanaka},
  \citenamefont {Saga},\ and\ \citenamefont {Mizokami}}]{Tanaka:85}%
  \BibitemOpen
  \bibfield  {author} {\bibinfo {author} {\bibfnamefont {K.}~\bibnamefont
  {Tanaka}}, \bibinfo {author} {\bibfnamefont {N.}~\bibnamefont {Saga}}, \ and\
  \bibinfo {author} {\bibfnamefont {H.}~\bibnamefont {Mizokami}},\ }\href
  {\doibase 10.1364/AO.24.001102} {\bibfield  {journal} {\bibinfo  {journal}
  {Appl. Opt.}\ }\textbf {\bibinfo {volume} {24}},\ \bibinfo {pages} {1102}
  (\bibinfo {year} {1985})}\BibitemShut {NoStop}%
\bibitem [{\citenamefont {Peyronel}\ \emph {et~al.}(2012)\citenamefont
  {Peyronel}, \citenamefont {Firstenberg}, \citenamefont {Liang}, \citenamefont
  {Hofferberth}, \citenamefont {Gorshkov}, \citenamefont {Pohl}, \citenamefont
  {Lukin},\ and\ \citenamefont {Vuleti{\'{c}}}}]{ref:Peyronel2012}%
  \BibitemOpen
  \bibfield  {author} {\bibinfo {author} {\bibfnamefont {T.}~\bibnamefont
  {Peyronel}}, \bibinfo {author} {\bibfnamefont {O.}~\bibnamefont
  {Firstenberg}}, \bibinfo {author} {\bibfnamefont {Q.-Y.}\ \bibnamefont
  {Liang}}, \bibinfo {author} {\bibfnamefont {S.}~\bibnamefont {Hofferberth}},
  \bibinfo {author} {\bibfnamefont {A.~V.}\ \bibnamefont {Gorshkov}}, \bibinfo
  {author} {\bibfnamefont {T.}~\bibnamefont {Pohl}}, \bibinfo {author}
  {\bibfnamefont {M.~D.}\ \bibnamefont {Lukin}}, \ and\ \bibinfo {author}
  {\bibfnamefont {V.}~\bibnamefont {Vuleti{\'{c}}}},\ }\href {\doibase
  10.1038/nature11361} {\bibfield  {journal} {\bibinfo  {journal} {Nature}\
  }\textbf {\bibinfo {volume} {488}},\ \bibinfo {pages} {57} (\bibinfo {year}
  {2012})}\BibitemShut {NoStop}%
\bibitem [{\citenamefont {Distante}\ \emph {et~al.}(2017)\citenamefont
  {Distante}, \citenamefont {Farrera}, \citenamefont {Padr{\'o}n-Brito},
  \citenamefont {Paredes-Barato}, \citenamefont {Heinze},\ and\ \citenamefont
  {de~Riedmatten}}]{ref:Distante2017}%
  \BibitemOpen
  \bibfield  {author} {\bibinfo {author} {\bibfnamefont {E.}~\bibnamefont
  {Distante}}, \bibinfo {author} {\bibfnamefont {P.}~\bibnamefont {Farrera}},
  \bibinfo {author} {\bibfnamefont {A.}~\bibnamefont {Padr{\'o}n-Brito}},
  \bibinfo {author} {\bibfnamefont {D.}~\bibnamefont {Paredes-Barato}},
  \bibinfo {author} {\bibfnamefont {G.}~\bibnamefont {Heinze}}, \ and\ \bibinfo
  {author} {\bibfnamefont {H.}~\bibnamefont {de~Riedmatten}},\ }\href {\doibase
  10.1038/ncomms14072} {\bibfield  {journal} {\bibinfo  {journal} {Nature
  Communications}\ }\textbf {\bibinfo {volume} {8}},\ \bibinfo {pages} {14072}
  (\bibinfo {year} {2017})}\BibitemShut {NoStop}%
\bibitem [{\citenamefont {Ryabtsev}\ \emph {et~al.}(2016)\citenamefont
  {Ryabtsev}, \citenamefont {Beterov}, \citenamefont
  {Tret{\textquotesingle}yakov}, \citenamefont {{\`{E}}ntin},\ and\
  \citenamefont {Yakshina}}]{ref:Ryabtsev_2016}%
  \BibitemOpen
  \bibfield  {author} {\bibinfo {author} {\bibfnamefont {I.~I.}\ \bibnamefont
  {Ryabtsev}}, \bibinfo {author} {\bibfnamefont {I.~I.}\ \bibnamefont
  {Beterov}}, \bibinfo {author} {\bibfnamefont {D.~B.}\ \bibnamefont
  {Tret{\textquotesingle}yakov}}, \bibinfo {author} {\bibfnamefont {V.~M.}\
  \bibnamefont {{\`{E}}ntin}}, \ and\ \bibinfo {author} {\bibfnamefont {E.~A.}\
  \bibnamefont {Yakshina}},\ }\href {\doibase 10.3367/ufne.0186.201602k.0206}
  {\bibfield  {journal} {\bibinfo  {journal} {Physics-Uspekhi}\ }\textbf
  {\bibinfo {volume} {59}},\ \bibinfo {pages} {196} (\bibinfo {year}
  {2016})}\BibitemShut {NoStop}%
\bibitem [{\citenamefont {Steck}(2010)}]{ref:Steck}%
  \BibitemOpen
  \bibfield  {author} {\bibinfo {author} {\bibfnamefont {D.}~\bibnamefont
  {Steck}},\ }\href@noop {} {\emph {\bibinfo {title} {Rubidium 87 D Line
  Data,}}}\ (\bibinfo  {publisher} {available online at
  http://steck.us/alkalidata},\ \bibinfo {year} {2010})\BibitemShut {NoStop}%
\bibitem [{\citenamefont {Asenjo-Garcia}\ \emph {et~al.}(2019)\citenamefont
  {Asenjo-Garcia}, \citenamefont {Kimble},\ and\ \citenamefont
  {Chang}}]{ref:AnaPNAS}%
  \BibitemOpen
  \bibfield  {author} {\bibinfo {author} {\bibfnamefont {A.}~\bibnamefont
  {Asenjo-Garcia}}, \bibinfo {author} {\bibfnamefont {H.~J.}\ \bibnamefont
  {Kimble}}, \ and\ \bibinfo {author} {\bibfnamefont {D.~E.}\ \bibnamefont
  {Chang}},\ }\href {\doibase 10.1073/pnas.1911467116} {\bibfield  {journal}
  {\bibinfo  {journal} {Proceedings of the National Academy of Sciences}\
  }\textbf {\bibinfo {volume} {116}},\ \bibinfo {pages} {25503} (\bibinfo
  {year} {2019})}\BibitemShut {NoStop}%
\bibitem [{\citenamefont {Balewski}\ \emph {et~al.}(2014)\citenamefont
  {Balewski}, \citenamefont {Krupp}, \citenamefont {Gaj}, \citenamefont
  {Hofferberth}, \citenamefont {Löw},\ and\ \citenamefont {Pfau}}]{ref:32}%
  \BibitemOpen
  \bibfield  {author} {\bibinfo {author} {\bibfnamefont {J.~B.}\ \bibnamefont
  {Balewski}}, \bibinfo {author} {\bibfnamefont {A.~T.}\ \bibnamefont {Krupp}},
  \bibinfo {author} {\bibfnamefont {A.}~\bibnamefont {Gaj}}, \bibinfo {author}
  {\bibfnamefont {S.}~\bibnamefont {Hofferberth}}, \bibinfo {author}
  {\bibfnamefont {R.}~\bibnamefont {Löw}}, \ and\ \bibinfo {author}
  {\bibfnamefont {T.}~\bibnamefont {Pfau}},\ }\href {\doibase
  10.1088/1367-2630/16/6/063012} {\bibfield  {journal} {\bibinfo  {journal}
  {New Journal of Physics}\ }\textbf {\bibinfo {volume} {16}},\ \bibinfo
  {pages} {063012} (\bibinfo {year} {2014})}\BibitemShut {NoStop}%
\bibitem [{\citenamefont {Browaeys}\ \emph {et~al.}(2016)\citenamefont
  {Browaeys}, \citenamefont {Barredo},\ and\ \citenamefont
  {Lahaye}}]{ref:Browaeys_2016}%
  \BibitemOpen
  \bibfield  {author} {\bibinfo {author} {\bibfnamefont {A.}~\bibnamefont
  {Browaeys}}, \bibinfo {author} {\bibfnamefont {D.}~\bibnamefont {Barredo}}, \
  and\ \bibinfo {author} {\bibfnamefont {T.}~\bibnamefont {Lahaye}},\ }\href
  {\doibase 10.1088/0953-4075/49/15/152001} {\bibfield  {journal} {\bibinfo
  {journal} {Journal of Physics B: Atomic, Molecular and Optical Physics}\
  }\textbf {\bibinfo {volume} {49}},\ \bibinfo {pages} {152001} (\bibinfo
  {year} {2016})}\BibitemShut {NoStop}%
\bibitem [{\citenamefont {B\'eguin}\ \emph {et~al.}(2013)\citenamefont
  {B\'eguin}, \citenamefont {Vernier}, \citenamefont {Chicireanu},
  \citenamefont {Lahaye},\ and\ \citenamefont {Browaeys}}]{ref:Ds}%
  \BibitemOpen
  \bibfield  {author} {\bibinfo {author} {\bibfnamefont {L.}~\bibnamefont
  {B\'eguin}}, \bibinfo {author} {\bibfnamefont {A.}~\bibnamefont {Vernier}},
  \bibinfo {author} {\bibfnamefont {R.}~\bibnamefont {Chicireanu}}, \bibinfo
  {author} {\bibfnamefont {T.}~\bibnamefont {Lahaye}}, \ and\ \bibinfo {author}
  {\bibfnamefont {A.}~\bibnamefont {Browaeys}},\ }\href {\doibase
  10.1103/PhysRevLett.110.263201} {\bibfield  {journal} {\bibinfo  {journal}
  {Phys. Rev. Lett.}\ }\textbf {\bibinfo {volume} {110}},\ \bibinfo {pages}
  {263201} (\bibinfo {year} {2013})}\BibitemShut {NoStop}%
\bibitem [{\citenamefont {Löw}\ \emph {et~al.}(2012)\citenamefont {Löw},
  \citenamefont {Weimer}, \citenamefont {Nipper}, \citenamefont {Balewski},
  \citenamefont {Butscher}, \citenamefont {Büchler},\ and\ \citenamefont
  {Pfau}}]{ref:experimental_data_Rydberg}%
  \BibitemOpen
  \bibfield  {author} {\bibinfo {author} {\bibfnamefont {R.}~\bibnamefont
  {Löw}}, \bibinfo {author} {\bibfnamefont {H.}~\bibnamefont {Weimer}},
  \bibinfo {author} {\bibfnamefont {J.}~\bibnamefont {Nipper}}, \bibinfo
  {author} {\bibfnamefont {J.~B.}\ \bibnamefont {Balewski}}, \bibinfo {author}
  {\bibfnamefont {B.}~\bibnamefont {Butscher}}, \bibinfo {author}
  {\bibfnamefont {H.~P.}\ \bibnamefont {Büchler}}, \ and\ \bibinfo {author}
  {\bibfnamefont {T.}~\bibnamefont {Pfau}},\ }\href {\doibase
  10.1088/0953-4075/45/11/113001} {\bibfield  {journal} {\bibinfo  {journal}
  {Journal of Physics B: Atomic, Molecular and Optical Physics}\ }\textbf
  {\bibinfo {volume} {45}},\ \bibinfo {pages} {113001} (\bibinfo {year}
  {2012})}\BibitemShut {NoStop}%
\bibitem [{\citenamefont {Hollerith}\ \emph {et~al.}(2019)\citenamefont
  {Hollerith}, \citenamefont {Zeiher}, \citenamefont {Rui}, \citenamefont
  {Rubio-Abadal}, \citenamefont {Walther}, \citenamefont {Pohl}, \citenamefont
  {Stamper-Kurn}, \citenamefont {Bloch},\ and\ \citenamefont
  {Gross}}]{ref:BlochExperiment}%
  \BibitemOpen
  \bibfield  {author} {\bibinfo {author} {\bibfnamefont {S.}~\bibnamefont
  {Hollerith}}, \bibinfo {author} {\bibfnamefont {J.}~\bibnamefont {Zeiher}},
  \bibinfo {author} {\bibfnamefont {J.}~\bibnamefont {Rui}}, \bibinfo {author}
  {\bibfnamefont {A.}~\bibnamefont {Rubio-Abadal}}, \bibinfo {author}
  {\bibfnamefont {V.}~\bibnamefont {Walther}}, \bibinfo {author} {\bibfnamefont
  {T.}~\bibnamefont {Pohl}}, \bibinfo {author} {\bibfnamefont {D.~M.}\
  \bibnamefont {Stamper-Kurn}}, \bibinfo {author} {\bibfnamefont
  {I.}~\bibnamefont {Bloch}}, \ and\ \bibinfo {author} {\bibfnamefont
  {C.}~\bibnamefont {Gross}},\ }\href {\doibase 10.1126/science.aaw4150}
  {\bibfield  {journal} {\bibinfo  {journal} {Science}\ }\textbf {\bibinfo
  {volume} {364}},\ \bibinfo {pages} {664} (\bibinfo {year}
  {2019})}\BibitemShut {NoStop}%
\end{thebibliography}%
\end{document}